\documentclass[11pt]{article}
\usepackage[american]{babel}
\usepackage{a4}
\usepackage{epsfig}
\usepackage{calc}
\usepackage{amsmath}
\chardef\letterchar=11
\chardef\otherchar=12
\renewcommand{\baselinestretch}{1.1}
\newdimen\fnindent
\makeatletter
\def\@makefntext#1{\noindent\hskip\fnindent\@makefnmark#1}
\makeatother
\begin{document}
\selectlanguage{american}
\sloppy
\newcommand{\gsim}     {\mbox{\raisebox{-0.4ex}{$\;\stackrel{>}{\scriptstyle 
\sim}\;$}}}
\newcommand{\lsim}     {\mbox{\raisebox{-0.4ex}{$\;\stackrel{<}{\scriptstyle 
\sim}\;$}}}
\newcommand{\tttld}    {\mbox{\raisebox{0.4ex}{\boldmath$\scriptscriptstyle\sim$}}}
\newcommand{\ep}       {\mbox{$e^{+}p$}}
\newcommand{\ncp}      {^{\rm NC}}
\newcommand{\dab}      {_{\rm \scriptscriptstyle DA}}
\newcommand{\born}      {_{\rm Born}}
\newcommand{\rad}      {_{\rm rad}}
\newcommand{\smod}      {^{\rm SM}}
\newcommand{\qq}       {\mbox{$Q^{2}$}}
\newcommand{\qqgt}     {\mbox{$Q^{2}>$ }}
\newcommand{\qqbt}     {\mbox{ $< Q^{2} <$ }}
\newcommand{\qqlt}     {\mbox{ $Q^{2}<$ }}
\newcommand{\qf}       {\mbox{$Q^{4}$}}
\newcommand{\qqc}      {\mbox{$Q^{2}_{c}$}}
\newcommand{\qqh}      {\mbox{$Q^{2}_{h}$}}
\newcommand{\qqda}     {\mbox{$Q^{2}\dab$}}
\newcommand{\xda}      {\mbox{$x\dab$}}
\newcommand{\yda}      {\mbox{$y\dab$}}
\newcommand{\gh}      {\mbox{$\gamma_h$}}
\newcommand{\Et}       {\mbox{$E_{T}$}}
\newcommand{\Pt}       {\mbox{$P_{T}$}}
\newcommand{\pt}       {\mbox{$p_{T}$}}
\newcommand{\ptgt}     {\mbox{$p_{T} >$ }}
\newcommand{\ptt}      {\mbox{$p_{T}^{2}$}}
\newcommand{\ptmin}    {\mbox{$p_{T}^{\rm min}$}}
\newcommand{\ptcut}    {\mbox{$p_{T}^{\rm cut}$}}
\newcommand{\kt}       {\mbox{$k_{T}$}}
\newcommand{\xgam}     {\mbox{$x_{\gamma}$}}
\newcommand{\xgamrec}  {\mbox{$x_{\gamma}^{\rm rec}$}}
\newcommand{\aetalt}   {\mbox{$\left| \eta \right| <$ }}
\newcommand{\ntrk}     {\mbox{$n_{\rm trk}$}}
\newcommand{\aventrk}  {\mbox{$\langle n_{\rm trk} \rangle$}}
\newcommand{\gev}      {\mbox{${\rm Ge\kern -0.1em V}$}}
\newcommand{\gevv}     {\mbox{${\rm Ge\kern -0.1em V}^{2}$}}
\newcommand{\sigqq}    {\mbox{$d\sigma/d\qq$}}
\newcommand{\sigxqq}   {\mbox{$d^{2}\sigma\ncp\born/dxd\qq$}}
\newcommand{\sigx}     {\mbox{$d\sigma/dx$}}
\newcommand{\sigy}     {\mbox{$d\sigma/dy$}}
\newcommand{\sigborn}  {\mbox{$\sigma_{\rm Born}$}}
\newcommand{\sigpt}    {\mbox{$d\sigma/d\pt$}}
\newcommand{\sigptt}   {\mbox{$d\sigma/d\ptt$}}
\newcommand{\sigeta}   {\mbox{$d\sigma/d\eta$}}
\newcommand{\sigetaptt}{\mbox{$d^{2}\sigma/d\eta d\ptt$}}
\newcommand{\chiz}     {\mbox{$\chi_{_{\rm Z}}$}}
\newcommand{\Mz}       {\mbox{$M_{\rm Z}$}}
\newcommand{\Mzq}      {\mbox{$M_{\rm Z}^{2}$}}
\newcommand{\thw}      {\mbox{$\theta_{\scriptscriptstyle \rm W}$}}
\newcommand{\sthw}     {\mbox{$\sin^2\!\thw$}}
\newcommand{\cthw}     {\mbox{$\cos^2\!\thw$}}
\newcommand{\alphas}   {\mbox{$\alpha_{\rm{s}}$}}
\newcommand{\onehalf}  {\mbox{${\textstyle{\frac{1}{2}}}$}}
%
\newlength{\alskip}
\settowidth{\alskip}{$^{  45}$}
\newcommand{\3}{\ss}
\newcommand{\lumi}     {\mbox{47.7 pb$^{-1}$}}
\newcommand{\lume}     {\mbox{$1.6\%$}}
\newcommand{\lw}[1]{\smash{\lower1.8ex\hbox{#1}}}
%
%
\title{\begin{flushright}{\large DESY--99--056}  \end{flushright}
\vspace*{2cm}
\LARGE \textbf{%
Measurement of High-\boldmath{\qq} Neutral-Current \\ 
\ep\ Deep Inelastic Scattering Cross-Sections \\
at HERA}}
\author{ZEUS Collaboration}
\date{ }
\maketitle
\noindent
\begin{abstract}
\noindent
The \ep\ neutral-current deep inelastic scattering differential
cross-sections \sigqq, for \qqgt 400~\gevv, \sigx\ and
\sigy, for \qqgt 400, 2500 and 10000~\gevv, have been measured with the ZEUS
detector at HERA. The data sample of \lumi\ was collected at a
center-of-mass energy of 300~\gev. 
The cross-section, \sigqq, falls by six
orders of magnitude between \qq~=~400 and 40000~\gevv. The
predictions of the Standard Model are in very good agreement with the
data. Complementing the observations of time-like $Z^0$ contributions to
fermion-antifermion annihilation, the data provide direct evidence for
the presence of $Z^0$ exchange in the space-like region explored by deep
inelastic scattering.
\end{abstract}

\pagestyle{plain}
\thispagestyle{empty}
\clearpage
\pagenumbering{Roman}
\begin{center}                                                                    
{                      \Large  The ZEUS Collaboration              }              
\end{center}                                                                      
  J.~Breitweg,                                                                    
  S.~Chekanov,                                                                    
  M.~Derrick,                                                                     
  D.~Krakauer,                                                                    
  S.~Magill,                                                                      
  B.~Musgrave,                                                                    
  A.~Pellegrino,                                                                  
  \mbox{J.~Repond,}  
  R.~Stanek,                                                                      
  R.~Yoshida\\                                                                    
 {\it Argonne National Laboratory, Argonne, IL, USA}~$^{p}$
% ...
\newpage                                                                          
% ...
$^{  36}$ partially supported by the Foundation for German-Russian 		  
Collaboration DFG-RFBR \\ \hspace*{3.5mm} (grant no. 436 RUS 113/248/3 		  
and no. 436 RUS 113/248/2)\\              					  
% ...
                                                           %
                                                           %
\newpage   
                                                           %
                                                           %
\begin{tabular}[h]{rp{14cm}}                                                      
$^{a}$ &  supported by the Natural Sciences and Engineering Research              
          Council of Canada (NSERC)  \\                                           
$^{b}$ &  supported by the FCAR of Qu\'ebec, Canada  \\                           
$^{c}$ &  supported by the German Federal Ministry for Education and              
          Science, Research and Technology (BMBF), under contract                 
          numbers 057BN19P, 057FR19P, 057HH19P, 057HH29P, 057SI75I \\             
$^{d}$ &  supported by the MINERVA Gesellschaft f\"ur Forschung GmbH, the         
German Israeli Foundation, and by the Israel Ministry of Science \\               
$^{e}$ &  supported by the German-Israeli Foundation, the Israel Science          
          Foundation, the U.S.-Israel Binational Science Foundation, and by       
          the Israel Ministry of Science \\                                       
$^{f}$ &  supported by the Italian National Institute for Nuclear Physics         
          (INFN) \\                                                               
$^{g}$ &  supported by the Japanese Ministry of Education, Science and            
          Culture (the Monbusho) and its grants for Scientific Research \\        
$^{h}$ &  supported by the Korean Ministry of Education and Korea Science         
          and Engineering Foundation  \\                                          
$^{i}$ &  supported by the Netherlands Foundation for Research on                 
          Matter (FOM) \\                                                         
$^{j}$ &  supported by the Polish State Committee for Scientific Research,        
          grant No. 115/E-343/SPUB/P03/154/98, 2P03B03216, 2P03B04616,            
          2P03B10412, 2P03B05315, 2P03B03517, and by the 
          German Federal Ministry of           
          Education and Science, Research and Technology (BMBF) \\                
$^{k}$ &  supported by the Polish State Committee for Scientific                  
          Research (grant No. 2P03B08614 and 2P03B06116) \\                       
$^{l}$ &  partially supported by the German Federal Ministry for                  
          Education and Science, Research and Technology (BMBF)  \\               
$^{m}$ &  supported by the Fund for Fundamental Research of Russian Ministry      
          for Science and Edu\-cation and by the German Federal Ministry for      
          Education and Science, Research and Technology (BMBF) \\                
$^{n}$ &  supported by the Spanish Ministry of Education                          
          and Science through funds provided by CICYT \\                          
$^{o}$ &  supported by the Particle Physics and                                   
          Astronomy Research Council \\                                           
$^{p}$ &  supported by the US Department of Energy \\                             
$^{q}$ &  supported by the US National Science Foundation \\                      
$^{r}$ &  partially supported by the British Council,                             
          ARC Project 0867.00 \\                                                  
$^{s}$ &  partially supported by the British Council,                             
          Collaborative Research Project, \\
       &  TOK/880/11/15
\end{tabular}
                                                           %
\newpage
\setcounter{page}{1}
\pagenumbering{arabic}
\renewcommand{\baselinestretch}{1.1}
\setlength{\parskip}{5pt plus 2pt minus 1pt}
\sloppy
\section{Introduction}

The HERA $ep$ collider has made possible the
exploration of deep inelastic scattering (DIS) in a new kinematic 
region, resulting in discoveries such as the rapid rise of the parton 
densities in the proton at low $x$-Bjorken~\cite{heralowx} and diffraction 
in DIS~\cite{heradiff}.  First measurements of charged and neutral-current (NC)
deep inelastic electron-proton scattering were made in a previously unexplored
region of large \qq~\cite{herahiq2}, where \qq\ is the negative square of 
the four-momentum transfer between the electron and proton.
Based on their 1994\,--1996 data, both the H1 and ZEUS collaborations have 
reported~\cite{h1highx,zhighx} more NC
events than expected from the Standard 
Model at high \qq\ and high $x$. These 
observations have prompted considerable discussion
in the particle-physics community as possible evidence for anomalies in the 
parton momentum distributions of the proton or of physics beyond the 
Standard Model.

This paper presents measurements at HERA of the
NC deep inelastic positron-proton scattering
differential cross-sections \sigqq\ for
\qqgt 400~\gevv\ and \sigx\ and \sigy\ for \qqgt 400, 2500, and
10000~\gevv, and their comparison to Standard Model predictions.  
This analysis is not optimized for the search for narrow lepton-hadron
resonances. The
measurements are based on \lumi\ of data collected by ZEUS from
1994\,--1997 during which HERA collided 27.5~\gev\ positrons with 
820~\gev\ protons, yielding a center-of-mass energy $\sqrt{s}~=~300$~\gev.
The highest \qq\ under study, 51200~\gevv, is much larger than the 
square of the $Z^0$-boson mass so that effects of $Z^0$ 
exchange are visible.

\section{Standard Model prediction}
\label{sc:standardmodel}

The electroweak Born--level NC DIS differential cross-section, \sigxqq,
for the reaction $e^{+} p \rightarrow e^{+} X$ can be expressed
\cite{xsec} as
\begin{equation}
  \frac{d^2\sigma\ncp\born(\ep)}{dxd\qq} = \frac{2\pi \alpha ^2}
    {x\qf}  \left[Y_+\,F_2\ncp(x,\qq)
    -  Y_-\,xF_3\ncp(x,\qq)  -y^2\,F_L\ncp(x,\qq)\right] ,
    \label{eq:born} 
\end{equation}
where $x$ is the Bjorken scaling variable, $\alpha(\qq\!=0) \simeq 1/137$
is the QED coupling constant, and $Y_{\pm} = 1 \pm (1-y)^2$ with 
$y=\qq/sx$.  The structure functions $F_2\ncp$ and $xF_3\ncp$ for 
longitudinally unpolarized beams may be described in leading order QCD 
as sums over the quark flavor $f=u,...,b$ of the product of electroweak
quark couplings and quark momentum distributions in the proton
\begin{alignat}{2}
  F_2\ncp  &=& \; \onehalf  &   \sum\limits_{f}\, {xq^+_f} \left[ 
    (V_f^L)^2+(V_f^R)^2+(A_f^L)^2+(A_f^R)^2 \right],\nonumber \\
  &&& \nonumber \\[-2.2em]
  &&& \\[-.8em]
  xF_3\ncp &=& &\sum\limits_{f}\, {xq^-_f} \left[ 
    V_f^LA_f^L-V_f^RA_f^R \right],\nonumber
\end{alignat}
where $xq^\pm_f=x q_f(x,\qq)\pm x {\bar q}_f(x,\qq)$ and
$x q_f$ ($x {\bar q}_f$) are the quark (anti-quark) momentum
distributions. In leading order QCD, we have $F_L\ncp = 0$.
The functions $V_f$ and $A_f$ can be written as
\begin{alignat}{2}
  V_{f}^{L,R} & = & \; e_{f} & -(v_{e} \pm a_{e})\,v_{f}\ 
  \chiz(\qq),\nonumber \\
  &&\nonumber \\[-2.2em]
  && \\[-0.8em]
  A_{f}^{L,R} & = & & -(v_{e} \pm a_{e})\,a_{f}\ 
  \chiz(\qq),\nonumber 
\end{alignat}
where the weak couplings, $a_i=T^3_i$ and $v_i=T^3_i - 2e_i\sthw$, are
functions of the weak isospin, $T^3_i = \frac{1}{2}$ ($-\frac{1}{2}$) for
$u,\nu$ ($d,e$), and the weak mixing angle, \thw; 
$e_i$ is the electric charge in units of the positron charge; and
\chiz\ is proportional to the ratio of $Z^0$-boson 
and photon propagators
\begin{equation}
\chiz =\frac{1}{4 \sthw \cthw}\frac{\qq}{\qq + \Mzq} .
\label{eq:chi} 
\end{equation}

All cross-section predictions in this paper are calculated using 
next-to-leading order (NLO) QCD
where the longitudinal structure function $F_L\ncp\neq 0$~\cite{fl}.
The contribution of $F_L\ncp$ to \sigxqq\ is 
predicted to be approximately $1.5\%$
averaged over the kinematic range considered in this paper.  However, 
in the region of small $x$ at the lower end of the \qq\ range the 
$F_L\ncp$ contribution to the cross-sections can be as large as $12\%$.

Uncertainties in the predicted cross-section arise from
three sources: electroweak parameters, electroweak radiative
corrections, and the parton momentum distributions including their higher
order QCD corrections. The electroweak parameters have been measured
to high precision by other experiments~\cite{electroweak,pdgcont} and
contribute less than $0.3\%$ uncertainty to the predicted cross-section
in the full kinematic range measured at HERA~\cite{ewhera}. Radiative
corrections for initial- and final-state radiation, vertex and
propagator corrections, and two-boson exchange have been calculated to
at least first order~\cite{heracles,hector}. Higher order corrections
for the kinematic region explored in this paper are expected to be
less than $1\%$~\cite{radcorr}. This leaves the parton momentum
distributions as the primary source of uncertainty in the predicted 
cross-section.

Parton momentum distributions have been determined by
several groups (GRV~\cite{grv}, MRS~\cite{mrs,mrst},
CTEQ~\cite{cteq,cteq4}) by parameterizing the distributions at some fixed
\qq\ and extrapolating the results to higher \qq\ using the DGLAP QCD
evolution equations~\cite{dglap}. The parameters are fitted to data
from lower energy fixed-target DIS experiments and from HERA, and,
in addition, to data measured at
the TEVATRON on lepton-pair production (Drell-Yan), direct
photon production, W production, and jet cross-sections. 
Note that the HERA data included in these parameterizations 
make their most significant contribution at $x < 10^{-2}$ and have 
relatively little influence on the predicted cross-sections used in this paper.
The sources of uncertainty in these fits can be divided
into two main groups: uncertainties in the measurements and
uncertainties in the fit itself. For the former, the statistical and
systematic uncertainties are available from each experiment.  For
the latter, uncertainties due to the QCD evolution were estimated by
varying the fit assumptions, such as the value of \alphas\ and higher twist.

Although the comprehensive parton momentum distribution fits from GRV, 
MRS, and CTEQ make extensive use of available data, they lack a 
complete estimate of uncertainties in the distributions.  
To obtain such an estimate, 
a NLO QCD fit was performed~\cite{fitMichiel}
to the DIS measurements of $F_2$ for 
proton and deuteron data from SLAC~\cite{fitSLAC}, BCDMS~\cite{fitBCDMS}, 
NMC~\cite{fitNMC}, E665~\cite{fitE665}, H1 and ZEUS~\cite{fitH1,94F2}, 
the measurements of $xF_3$ from neutrino measurements by 
CCFR~\cite{fitCCFR}, and the $\bar d - \bar u$ data from E866~\cite{e866}. 
Included in 
the fit were statistical and correlated systematic errors from each 
experiment.  Also considered were the effects of a change of 
$\alphas(\Mzq)$ from 0.113 to 0.123, a $50\%$ variation in the 
strange quark content of the proton, a variation of the factorization and
renormalization scales $\mu_{f,r}$ in the range $Q^2/2 < \mu_{f,r}^2 < 2Q^2$,
and of corrections for nuclear effects, 
all of which produced uncertainties of typically less than $1\%$. 
The results are in good agreement
with the MRST~\cite{mrst} and CTEQ4~\cite{cteq4} fits, and differences 
are typically smaller than the extracted uncertainties.
The fit yields uncertainties in the cross-section \sigqq\ of 
approximately $2.5\%$ for \qq~=~400~\gevv\ and $6\%$ 
at the highest \qq\ under study.

Other uncertainties were also investigated.  For example, charm-threshold
effects, calculated from three different models of charm evolution in
the proton as supplied by CTEQ~\cite{cteq4charm}, produced cross-sections 
that differed by less than $3\%$. 
An analysis of the stability of perturbative calculations of the 
production of bottom- and charm-quarks at HERA~\cite{grshq} showed 
negligible effects. The MRST~\cite{mrst} fit
incorporating the latest prompt photon data at high $x$ from
E706~\cite{e706} and $\bar d - \bar u$ data from E866~\cite{e866} 
produced cross-sections 
that were lower than those calculated using CTEQ4 by $4\%$ at 
\qq\ =~400~\gevv, increasing to a maximum of $8\%$ at \qq\ = 10000~\gevv.  
The CTEQ4HJ~\cite{cteq4} fit, specifically tuned to reproduce the jet 
high transverse
energy cross-section reported by CDF~\cite{cdf}, produced changes in
the cross-section of less than $2\%$ except at the highest \qq\ 
=~50000~\gevv, where it yielded an increase of $6\%$.
The CTEQ5~\cite{cteq5} fit incorporating more data than the CTEQ4 fit,
in particular introducing an improved handling of $d/u$ and $\bar d-\bar u$
using E866 data and a measurement of charge asymmetry in W-production at the
TEVATRON~\cite{cdfW}, gave cross-sections that were higher by $3\%$ at
\qq\ =~400~\gevv\ and lower by $2\%$ at \qq\ =~10000~\gevv.

We conclude from these studies that the parton densities give a total
uncertainty on the Standard Model prediction of the NC DIS
differential cross-section \sigqq\ of $4\%$ for \qq\ = 400~\gevv\ increasing
to $8\%$ at the highest \qq\ in the $x$--range covered by this measurement.
In the following, uncertainties in the parton momentum distributions
are taken from the ZEUS\ NLO\ QCD\ fit~\cite{fitMichiel}.

\section{The ZEUS experiment}
\label{sc:zeus}

ZEUS~\cite{zeus} is a multipurpose magnetic detector designed 
to measure $ep$ interactions at HERA.  The primary components used 
for this analysis are the compensating uranium-scintillator 
calorimeter (CAL), the central tracking detector (CTD), and the luminosity 
detector.

The ZEUS coordinate system is right-handed with the $Z$ axis pointing in the
direction of the proton beam (forward) and the $X$ axis pointing horizontally
toward the center of HERA. The polar angle $\theta$ is zero in the 
$Z$ direction.

The CAL~\cite{cal} covers $99.7\%$ of the total solid angle. 
It is divided into three parts
with a corresponding division in 
$\theta$ as viewed from the nominal interaction point: 
forward (FCAL, $2.6^\circ < \theta < 36.7^\circ$), barrel
(BCAL, $36.7^\circ < \theta < 129.1^\circ$), and rear (RCAL,
$129.1^\circ < \theta < 176.2^\circ$). Each section is subdivided into
towers which subtend solid angles between 0.006 and 0.04 steradian.
Each tower is longitudinally segmented into an electromagnetic (EMC) and 
one (RCAL) or two (FCAL, BCAL) hadronic sections (HAC). The electromagnetic
section of each tower is further subdivided transversely into two
(RCAL) or four (BCAL, FCAL) cells. Under test beam conditions the
calorimeter resolutions are $\sigma/E = 18\%/\sqrt{E\,(\gev)}$ for
electrons and $\sigma/E = 35\%/\sqrt{E\,(\gev)}$ for hadrons. The
calorimeter has a time resolution of better than 1~ns for energy deposits
above 4.5~\gev.

A presampler detector is mounted in front of FCAL and RCAL. It
consists of scintillator tiles matching the calorimeter towers and
measures signals from particle showers in the material between the
interaction point and the calorimeter.

Tracking information is provided by the CTD~\cite{ctd} operating in 
a 1.43 T solenoidal magnetic field.  The interaction vertex
is measured with a typical resolution along (transverse to)
the beam direction of 0.4~(0.1)~cm.
The CTD is used to reconstruct the momenta of 
tracks in the polar angle region $15^\circ < \theta < 164^\circ$.  
The transverse momentum ($p_t$) resolution for 
full-length tracks can be parameterized 
as $\sigma(p_t)/p_t=0.0058\ p_t \oplus 0.0065 \oplus 0.0014/p_t$,
with $p_t$ in GeV.

The luminosity is measured using the Bethe-Heitler reaction $ep
\rightarrow ep\gamma$~\cite{lumi}. The resulting small angle
energetic photons are measured by the luminosity monitor, a
lead-scintillator calorimeter placed in the HERA tunnel 107 m from the
interaction point in the positron beam direction.

\section{Monte Carlo simulation}
\label{sc:mc}

Monte Carlo simulations (MC) are used to determine the efficiency for 
selecting events, to determine the accuracy of kinematic 
reconstruction, to estimate the background rate, and to extrapolate
the measured cross-sections 
to the full kinematic phase space.  A sufficient number of 
events is generated to ensure that errors from MC statistics can be neglected.
The MC samples are normalized to the total
integrated luminosity of the data.

The ZEUS detector response is simulated with a program based on 
GEANT~\cite{geant}.  The generated events are passed through 
the simulated detector, subjected to the same trigger requirements as 
the data, and processed by the same reconstruction programs.

The vertex distribution is a crucial input to the MC
as this is necessary to estimate the event selection efficiency. The
latter is strongly correlated with the $Z$-coordinate of the
event vertex, as the reconstructed values of \qq, \Et\ and other 
quantities depend on the measured vertex position. 
For the 1995 to 1997 data sets, the underlying distribution 
of the $Z$-coordinate of the event vertex is
determined using a minimum-bias sample of low \qq\ neutral-current DIS
events. For 1994 data this method is compared to the estimate from a
special minimum-bias soft photoproduction trigger, where very good
agreement in shape is found. 
The uncertainty in the shape of the vertex $Z$-distribution
is related to the fraction of $ep$ collisions from RF buckets adjacent to that
containing the main proton beam. 
The effect of this uncertainty on the normalization of the data sample in the 
vertex range -50 to {\small +}50~cm is estimated to be less than $0.4\%$.

NC DIS events including radiative effects are simulated 
using the HERACLES 4.5.2~\cite{heracles} program with the 
DJANGO6 2.4~\cite{django} interface to the hadronization programs.  
In HERACLES, corrections for initial- and final-state radiation, vertex and 
propagator corrections, and two-boson exchange are included.  
The QCD cascade and the hadronic final state are simulated
using the color-dipole model of ARIADNE 4.08~\cite{ariadne} and, as a
systematic check, with the MEPS model of LEPTO 6.5~\cite{meps}. Both
programs use the Lund string model of JETSET 7.4~\cite{jetset} 
for the hadronization.

Photoproduction background is estimated using events 
simulated with HERWIG~\cite{herwig}. In 
addition, a large sample of prompt photon events ($ep\rightarrow e\gamma X$), 
is generated with HERWIG.

\section{Reconstruction of kinematic variables}
\label{sc:reco}

\subsection{Event characteristics}
\label{sc:evtchar}

Neutral-current DIS at the high-\qq\ values 
discussed here produces striking events, relatively easy to 
distinguish from the potentially large backgrounds of quasi-real photoproduction 
(\qq $\sim$ 0) and beam-gas interactions.  The events are 
characterized by a high-energy isolated positron in the detector.  
For \qqgt 400~\gevv, most of 
the positrons have an energy near the positron beam energy and
are restricted to a polar angle below
$140^\circ$.  As \qq\ increases, the positrons are produced with higher 
energies, up to several hundred GeV, and at smaller polar angles.

The variables $\delta$, \Pt\ and \Et\ are used for event selection. 
These are defined as
\begin{eqnarray}
\delta &=& \sum\limits_{i} ( E_i - E_i \cos \theta_{i} ) 
= \sum\limits_{i} (E-p_z)_{i},
\label{eq:delta}
\end{eqnarray}
where the sum runs over all calorimeter energy deposits $E_i$ 
(uncorrected in the trigger, but corrected energies in the
offline analysis as discussed below)
with polar angle $\theta_i$.  
At the generator level $\delta = 55$~GeV, i.e.\ twice the positron 
beam energy, which follows from
energy-momentum conservation. Undetected particles which escape
through the forward beam hole give a negligible change in $\delta$ 
while particle loss
through the the rear beam hole, e.g.\ from initial state bremsstrahlung
or for photoproduction background, 
can lead to a substantial reduction of $\delta$.
The net transverse momentum, \Pt, and the transverse energy, \Et, 
are defined by
\begin{alignat}{2}
P_T^2  & = & P_x^2 + P_y^2                = &
  \left( \sum\limits_{i} E_i \sin \theta_i \cos \phi_i \right)^2+
\left( \sum\limits_{i} E_i \sin \theta_i \sin \phi_i \right)^2,
  \label{eq:pt}\\ 
E_T    & = & \sum\limits_{i} E_i \sin \theta_i, &\label{eq:et}
\end{alignat}
where $\phi_i$ is the azimuthal angle and the sums run, as above, over
all energy deposits in the calorimeter.  
High-\qq\ events in which the positron strikes
the BCAL or FCAL (\qqgt 1000~\gevv) are characterized by large \Et.

In the determination of the DIS kinematics, the CAL energy deposits
are separated into those associated with the identified  scattered 
positron, and all other energy deposits.  The sum of the latter is referred to 
as the hadronic energy. 

\subsection{The double-angle method}

\qq, $x$, and $y$ are measured using the double-angle 
meth\-od~\cite{recmethod}
\begin{alignat}{2}
\qqda &=& \, 4E_e^{2} & \frac{\sin\gamma_h(1+\cos\theta_e)}
              {\sin\gamma_h + \sin\theta_e - \sin(\gamma_h+\theta_e)}, \\
\xda  &=& \, \frac{E_e}{E_p} &
           \frac{\sin\gamma_h + \sin\theta_e + \sin(\gamma_h+\theta_e)}
            {\sin\gamma_h + \sin\theta_e - \sin(\gamma_h+\theta_e)}, \\
\yda  &=& & \frac{\sin\theta_e(1-\cos\gamma_h)} 
           {\sin\gamma_h + \sin\theta_e - \sin(\gamma_h+\theta_e)},
\end{alignat}
where $E_e$ ($E_p$) is the energy of the positron (proton) beam, 
$\theta_e$ is the polar angle
of the scattered positron, and \gh, in the na\"{\i}ve quark
parton model picture of DIS, is the polar angle of the struck quark.
The determination of the angles $\theta_e$ and \gh\ is discussed
in Sect.~\ref{sc:recofinder} and~\ref{sc:recogamma}, respectively.

The double-angle method is essentially
insensitive to errors in the absolute energy scale of the calorimeter.  
However it is sensitive to QED radiation and 
an accurate simulation of the hadronic final state is necessary.
At \qqgt 400~\gevv\ the hadronic final state 
is sufficiently energetic that model uncertainties of fragmentation and the 
treatment of the proton remnant are less important than at lower \qq.

To validate the performance of the double-angle meth\-od, the
reconstructed kinematic variables of MC events are compared 
to the true hadron variables \qq, $x$ and $y$
as defined by the four-momentum transfer $q$ to the hadronic system
\begin{equation}
\qq = -q^2, {\hskip .2cm} x = \qq / (2 p \cdot q ),
{\hskip .2cm} y = \qq/(x s)
\label{eq:hadronvar}
\end{equation}
where $p$ is the four-momentum of the initial proton and $s = 4 E_p E_e$.  
The resolution in the kinematic variables is determined accordingly 
(see Sect.~\ref{sc:unfold}) and demonstrates (not shown here) 
that the double-angle method performs 
better than other methods for \qqgt 400~\gevv.

\subsection{Positron identification}
\label{sc:recofinder}

A key signature of the events under study is the presence of an 
isolated high-energy positron. In order to identify and reconstruct the
positron, while rejecting events in which other final state particles mimic 
it, an algorithm is used which combines calorimeter and CTD
information.

In a first step, calorimeter ``clusters'' are formed by grouping all CAL cells
with energy depositions into cones relative to the event interaction point, 
which are centered
around cells with a local energy maximum~\cite{ZUFO}. All clusters are
treated as positron candidates. The cluster energy, $E_{{\rm clu}}$, is
the sum of the cell energies belonging to the cluster.  
The center of each cluster is calculated by weighting each tower
member by a logarithmic function of its energy.  
The cluster angle, $\theta_{{\rm clu}}$, is set equal to the
polar angle obtained from the center position of the cluster and 
the event vertex obtained from the tracks
measured with the CTD.  For clusters with polar 
angle\footnote{We do not consider candidates with 
$\theta_{{\rm clu}}>164^{\circ}$
(which are also beyond the CTD acceptance limit), since they
correspond to $Q^2$ values below the range of this analysis.} within
the CTD acceptance ($\theta_{{\rm clu}} >17.2^{\circ}$), a matching
track is required. A track is considered to match if the distance of
closest approach (DCA) between the extrapolation of the track
into the calorimeter and the position of the cluster center is less than 
$10$~cm, where the r.m.s.\ resolution in the DCA is 1.8~cm.

In the second step, several quantities, $\xi _i$, are calculated for
each positron candidate: the fraction of the cluster energy in the
hadronic sections of the calorimeter; the parameters related to
lateral energy profiles; and the total energy
in all calorimeter cells not associated with the cluster but lying
within a cone in pseudorapidity, $\eta = -\ln(\tan(\theta/2))$,
and azimuthal angle, $\Phi$, of radius 
$R=\sqrt{(\Delta\eta)^2+(\Delta\Phi)^2}=0.8$, 
centered on the cluster.  If a
matching track is present, the polar and azimuthal
angle differences between the track and the cluster position, and the
quantity $1/E_{{\rm clu}}-1/{P_{{\rm trk}}} $, where ${P_{{\rm trk}}}$
is the track momentum, are evaluated.

Finally, for each $\xi _i$ a probability $P(\xi_i)$ is derived,
designed to be uniformly distributed between zero and one for positrons.
Candidates are accepted as positrons if the product of the $P(\xi_i)$
exceeds a threshold determined from MC studies.  Since the CAL energy 
resolution is better than that of the CTD for tracks above 10~\gev, for 
accepted candidates, the positron energy, $E_e^\prime$, is set equal to the
cluster energy, $E_{{\rm clu}}$, comprising typically six cells.
The positron angle, $\theta_e$, is determined 
from the associated track if the positron cluster
is within the CTD acceptance, and otherwise set equal to $\theta_{\rm clu}$.
The resolution in $\theta_e$ is typically~0.1$^\circ$.
Sect.~\ref{sc:selec:offline} contains further selection criteria 
applied to the positron candidates.

\subsection{Reconstruction of the hadronic final state}
\label{sc:recogamma}

Energy depositions of the hadronic final state are used to evaluate the 
angle \gh. For this purpose calorimeter clusters are used as just described.
The angle \gh\ is then calculated for the event according to
\begin{equation}
\cos\gamma_h = \frac{P_{T,h}^2 - \delta^2_h}{P_{T,h}^2 + \delta^2_h} ,
\label{eq:gammah}
\end{equation}
where $P_{T,h}$ and $\delta_h$ are calculated using~(\ref{eq:delta}) 
and~(\ref{eq:pt}) with sums running over the
calorimeter clusters in the same manner as \Pt\ and $\delta$, but excluding
the positron.

\begin{figure} [p]
\centerline{\epsfig{file=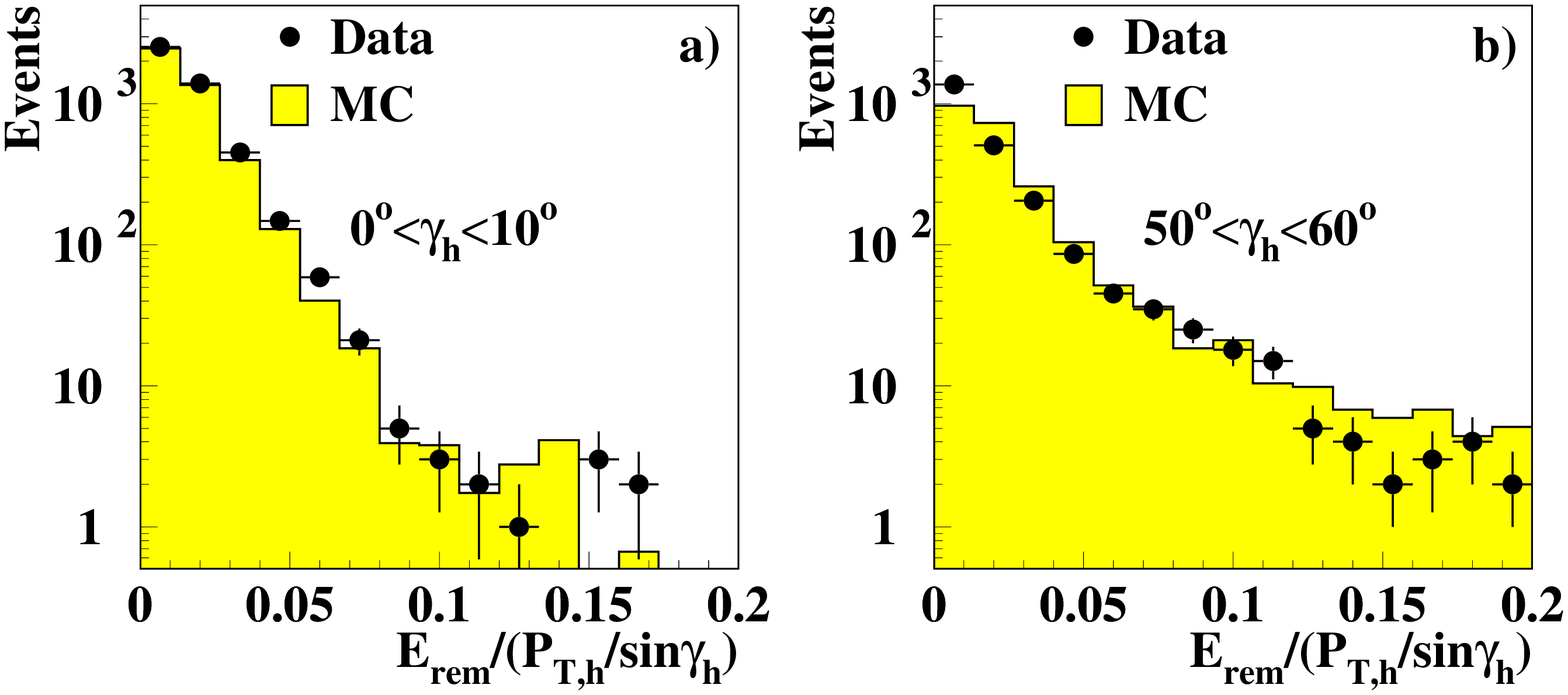,width=\linewidth}}
\centerline{\epsfig{file=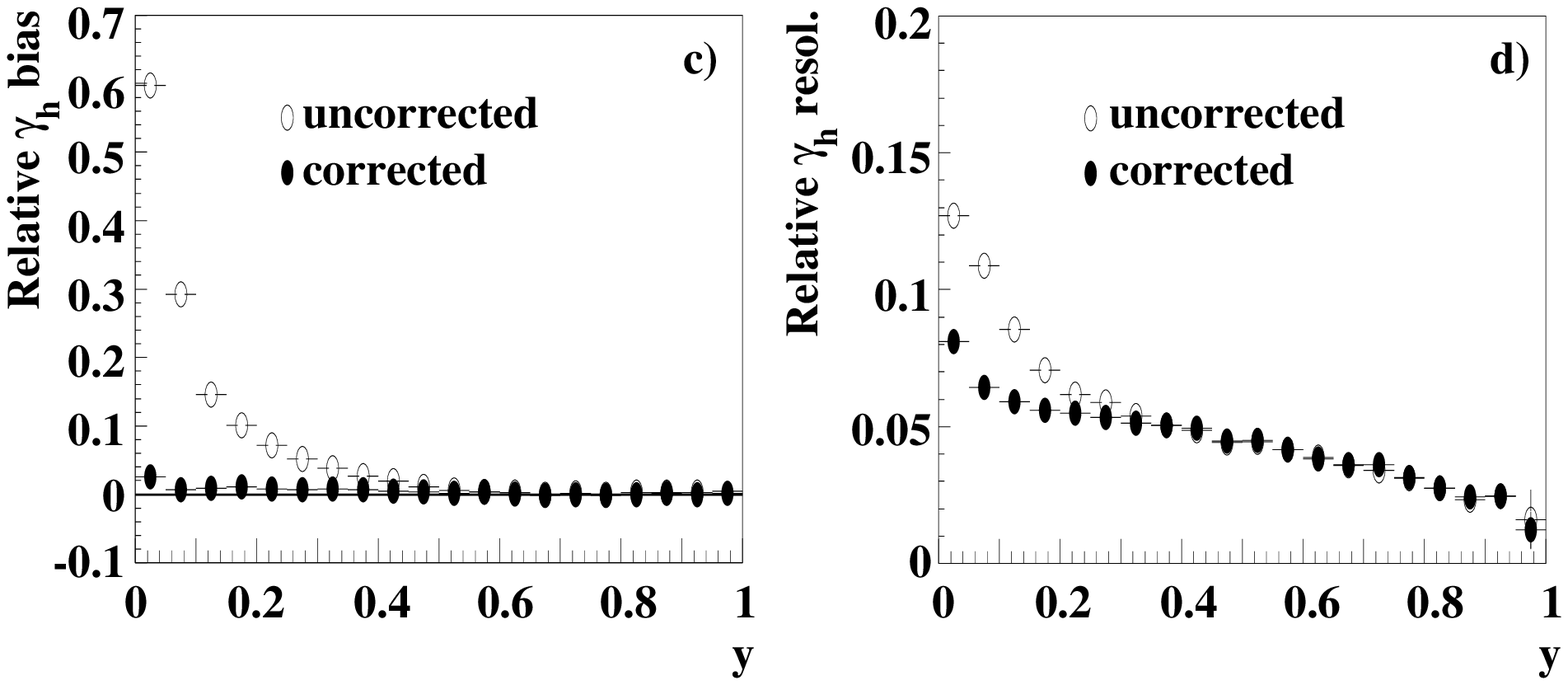,width=\linewidth}}
\caption[Fig??]{The effect of the correction used for the reconstruction 
  of the angle \gh\ as described in Sect.~\ref{sc:recogamma}. 
  (a) and (b): the ratio of total energy $E_{\rm rem}$ removed by
  the correction and the hadronic transverse
  momentum $P_{T,h}$ divided by $\sin \gh$ for two different
  ranges in \gh. The improvement in the reconstruction of \gh\ 
  by applying the correction is shown for the relative bias in \gh\ (c) and
  the relative resolution in \gh\ (d), where relative refers to the
  normalization to the true $\gamma$; both are shown as a function of
  the true $y$ as defined in~(\ref{eq:hadronvar}).  
  The result with (without) correction as obtained from a NC MC
  is shown as filled (unfilled) symbols.}
\label{fg:gamma}
\end{figure}

MC studies of the calorimeter response indicate that the
uncorrected \gh\ calculated with~(\ref{eq:gammah}) is biased
by redirected hadronic energy from interactions in material
between the primary vertex and the calorimeter or by
backsplash from the calorimeter (albedo)\footnote{High energy hadrons 
interacting in a calorimeter can produce with a non-negligible
probability particles at large angles with respect
to the direction of the main shower.
Some of these particles travel backwards and generate energy
deposits far away from their primary source in
the calorimeter which is referred to as backsplash.}.  
To minimize this bias, clusters with energy below 3~\gev\ 
and with polar angles larger than $\gamma_{\rm max}$ are
removed. The value of $\gamma_{\rm max}$, which is a function
of \gh, is derived from a NC MC
sample by requiring that less than $1\%$ of the clusters not related to 
the above effects be removed. This yields a reconstruction
of \gh\ closest to the true value as given by the MC.
The agreement of the distribution of removed
energies for different \gh-ranges between data and MC
is reasonable as shown in Fig.~\ref{fg:gamma}(a) and (b). After this first pass
of cluster removal the value of \gh\ is re-calculated and the 
procedure is repeated until it converges, typically
after two or three passes.  Removing calorimeter clusters 
in this manner substantially improves the resolution and bias of the 
double-angle variables for small values of \gh\ (corresponding to small
values of $y$) as shown from the ARIADNE MC
in Fig.~\ref{fg:gamma}(c) and (d) and leaves them largely
unchanged for large values of \gh. A similar result is achieved
with the LEPTO MC (not shown here).

\subsection{Energy correction and calorimeter performance}
\label{sc:cperf}
{\bf Correction for energy loss: }
All energies of clusters from both the positron and the
hadronic final state are
corrected for energy loss in the material between the interaction point and the
calorimeter.  All hadronic clusters and the positron clusters
in BCAL and FCAL are corrected based on the material maps implemented
in the detector simulation package. 
The presampler is used to correct positrons entering the RCAL.
This correction is based on the measured positron
energy, the amount of material in front of the RCAL and the presampler
signal.

\begin{figure} [p]
\centerline{\epsfig{file=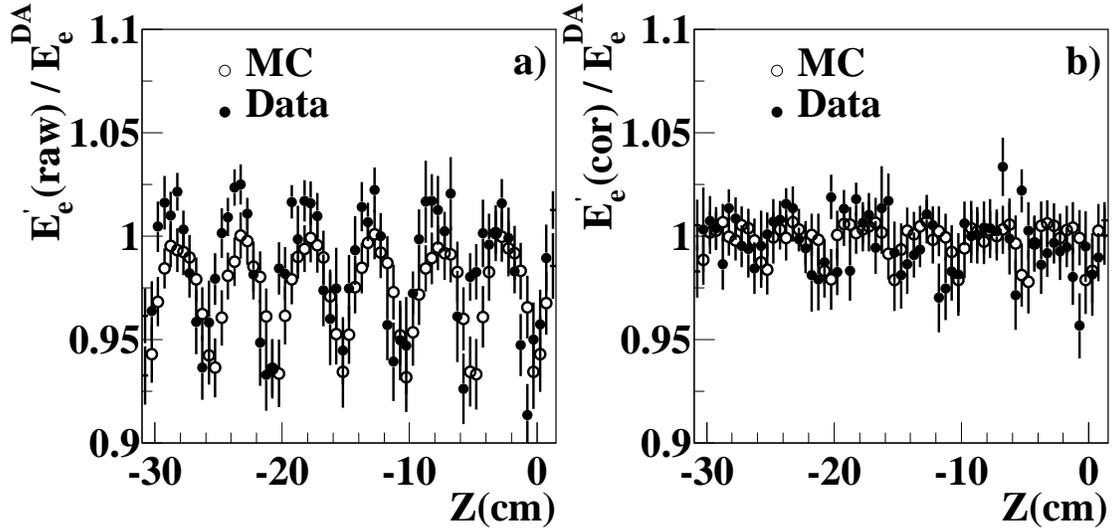,width=\linewidth}}
\caption[Fig??]{(a) the ratio of the positron energy $E_e^\prime({\rm raw})$ 
  measured in BCAL 
  to that calculated from double-angle variables, $E_e^{\rm DA}$, as a 
  function of the $Z$-position of the positron in BCAL. The dips coincide
  with the cell boundaries in BCAL. (b)
  the same after applying a non-uniformity correction as described in
  Sect.~\ref{sc:cperf}, yielding $E_e^\prime({\rm cor})$. 
  Open circles show MC and dots show data.}
\label{fg:nonunfty}
\end{figure}

{\bf Non-uniformity correction for positrons: }
In Fig.~\ref{fg:nonunfty}(a) the ratio of the positron energy corrected
as described above to the double-angle prediction is shown. 
At boundaries of calorimeter cells and modules 
there are dips in the measured energy response.
This effect is larger in data than in MC.
To account for this effect a correction is obtained from
fits to the non-uniformity patterns in the BCAL.
After correction, the data are well reproduced by
the MC as shown in Fig.~\ref{fg:nonunfty}(b).
A similar correction is used for geometrical effects in the azimuthal angle.

{\bf Calorimeter energy scale: }
The uncertainty in the energy scale of the scattered positron
is determined after applying the corrections described above.
For scattered positrons in low $y$ 
DIS events ($\theta_e\gsim135^\circ$ due to the requirement $y<0.04$),
the energy is strongly correlated with the scattering angle, and 
a comparison of the
predicted energy to the measured energy in the calorimeter is made.
This method is described in detail elsewhere~\cite{94F2}.
In the range of $30^\circ < \theta_e < 150^\circ$, the momentum of the
positrons can be determined by the CTD.  
The average track momentum minus calorimeter energy is used as an independent
check of the energy measured in the CAL for energies up to 30~\gev. 
For positrons with energies above 30~\gev, or those 
scattered to extreme forward directions, a comparison
of the energy predicted from double-angle variables
to the measured energy is made. In kinematic regions
where the other methods can be used to check the CAL energy scale, 
the double-angle results are in 
agreement with the other methods to better than $0.5\%$.
As a result of these studies,
the uncertainty in the value of the energy of the scattered positron in the
RCAL is $2\%$ at 10~\gev, decreasing linearly to $1\%$ at 27.5~\gev\ and above,
$1\%$ in the BCAL and $3\%$ in the FCAL.
  
\begin{figure}[p]
\centerline{\epsfig{file=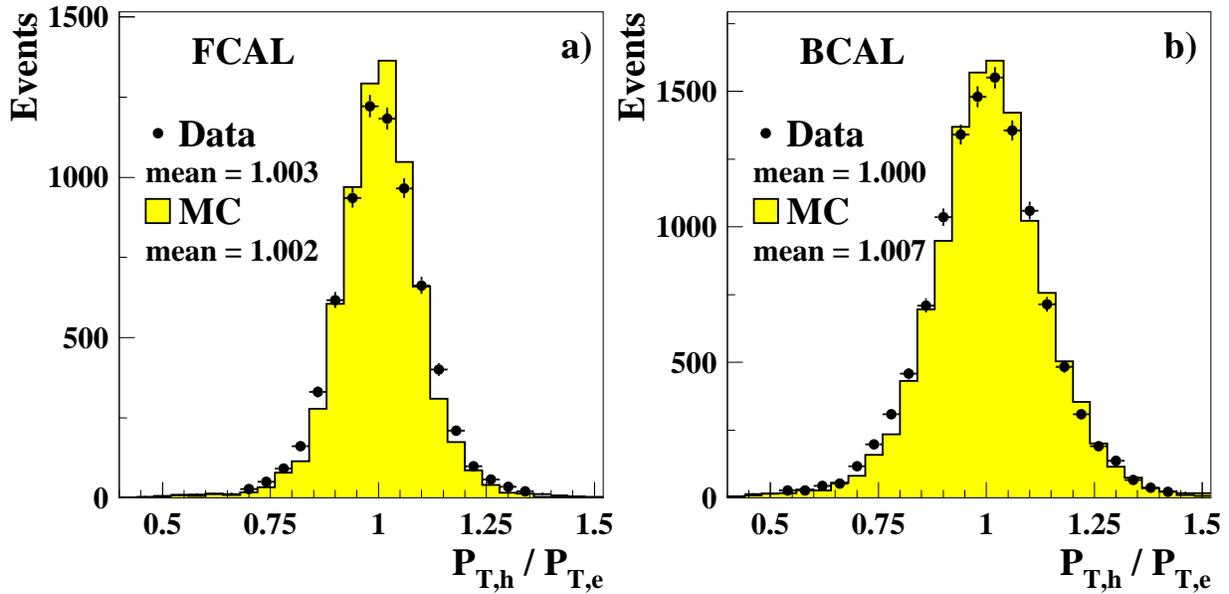,width=\linewidth}}
\caption[Fig??]{A comparison of data (dots) and MC
  (histogram) for the ratio of the transverse momentum of the
  hadronic final state, $P_{T,h}$, and the positron, $P_{T,e}$, for 
  $P_{T,h}>20$~GeV and \gh\ pointing into the FCAL (a) and into 
  the BCAL (b).}
\label{fg:escale}
\end{figure}

The uncertainty in the scale of the hadronic energy has been studied.  
After applying the corrections to the
energy described above, the ratio of the hadronic
transverse momentum, $P_{T,h}$, to the transverse momentum $P_{T,e}$ carried
by the positron found in the BCAL, is examined event by event. 
For transverse momenta above 20~\gev, the
ratio is $1.0$ to within $0.5\%$ both for data and MC, as expected
from momentum conservation (see Fig.~\ref{fg:escale}).
For lower transverse energies, the ratio is below $1.0$ by up to 
several percent due to hadronic energy loss. The comparisons 
between MC predictions of the transverse momentum ratio and the data agree
to within 1-$2\%$.
A comparison with corresponding results using the ratio of $P_{T,h}$ to the
transverse momentum predicted from double-angle variables
also shows good agreement. As a result of these studies,
the uncertainty in the hadronic energy scale is determined to be $2\%$ 
in the FCAL and BCAL, of which $1\%$ comes from the uncertainty of the positron
measurement.  The dominant remaining uncertainty comes from comparison
of MC results for simulations of hadronic final states
using ARIADNE~\cite{ariadne} and HERWIG~\cite{herwig}. 
The uncertainty is $3\%$ in the RCAL
where the hadronic energy is lower than in FCAL and BCAL
due to kinematic constraints. 

{\bf Positron energy resolution: }
After all corrections, the resolution in positron energy is evaluated
by comparing the width of the distribution of the
ratio of corrected energy to the value
predicted from the double-angle method. Since the resolution is smaller in
the MC than in data, the MC energies in BCAL and RCAL are smeared
accordingly. For RCAL a constant relative smearing of $1.7\%$
is applied while for BCAL the dependence of the resolution on the
calorimeter non-uniformity is taken into account. 

{\bf Hadronic energy resolution: }
A similar method, based on $P_{T,h}$, is used to determine the 
resolution in hadronic energy.
Again, the resolution is found to be smaller in the MC than in data.
A smearing is applied accordingly to clusters in MC for all calorimeter parts.

\subsection{Detector alignment}

The polar and azimuthal angles of the scattered positron can be
measured with the tracking detectors as well as with the calorimeter.
Comparisons of the position of 
calorimeter non-uniformities resulting from the cell 
structure (see Sect.~\ref{sc:cperf}) and
the extrapolated positron position from the CTD
allow the BCAL and RCAL to be aligned in $Z$ with respect to the CTD
to 1 and to 2~mm, respectively, and to be aligned in 
transverse direction to 2 and to 1~mm, respectively.

\section{Event selection}
\label{sc:selec}

\subsection{Trigger}
\label{sc:selec:trigger}

The ZEUS trigger operates at three levels~\cite{zeus}.  For the
portions of the trigger relevant to this analysis, the requirements
were strictest during 1997 and are described here.  The first-level
trigger requires a total electromagnetic energy of at least 
3.4~\gev\ in the RCAL
or 4.8~\gev\ in the BCAL, or $E_T^{\prime\prime} >$ 30~\gev, 
where $E_T^{\prime\prime}$ is the total transverse energy excluding 
the two rings of calorimeter towers nearest to the forward beampipe.  
The $E_T^{\prime\prime}$ requirement is designed
to tag high-\qq\ events by their large \Et\ while rejecting beam-gas
background having large energy deposits at low polar angles.  The
major requirement at the second-level trigger is $\delta + 2E_\gamma >$
29~\gev, where $E_\gamma$ is the energy measured in the luminosity
monitor.  This requirement suppresses photoproduction. 
Backgrounds are further reduced at the second-level trigger by
removing events with calorimeter timing inconsistent with an $ep$
interaction.  For the third-level trigger, events are reconstructed
on a computer farm, and the requirements are similar to the offline
cuts described below, though looser and using a simpler and
generally more efficient (but less pure) positron finder.

\subsection{Offline selection}
\label{sc:selec:offline}

The following criteria are applied offline (see
also Sect.~\ref{sc:reco} and~\cite{zhighx}):
\begin{itemize}
\item To ensure that event quantities can be accurately determined, a
  reconstructed vertex with $-50 < Z < 50$ cm is required, a range
  consistent with the $ep$ interaction region.
\item To suppress photoproduction events where the scattered positron
  escapes through the beam hole in the RCAL, $\delta$ is required to
  be greater than 38~\gev. This cut also reduces the number of events
  with initial-state QED radiation. The requirement 
  $\delta<65$~\gev\ removes cosmic ray background.
\item Positrons are identified based on calorimeter cluster quantities
  and tracking.
\begin{itemize}
\item To ensure high purity, the positron is required to have an
  energy of at least 10~\gev; in this case the identification efficiency 
  exceeds $96\%$, as shown by MC studies.
\item To reduce background, isolated positrons are selected by
  requiring no more than 5~\gev\ in calorimeter cells not associated
  with the scattered positron in an $\eta-\Phi$ cone of radius 0.8
  centered on the positron. 
\item In addition, each positron with $\theta_e
  > 17.2^\circ$ must be matched to a charged track of
  at least 5~\gev\ momentum.
\item For positrons beyond the forward tracking acceptance ($\theta_e
  < 17.2^\circ$), the tracking requirement in the positron selection is
  replaced by a cut on the transverse momentum of the
  positron $P_{T,e} > 30$~\gev\ and by the requirement $\delta > 44$~\gev.
\item A fiducial volume cut is applied to the positron position. 
  This excludes 
  the upper part of the central RCAL area ($20\times 80$~cm$^2$) 
  occluded by the cryogenic supply for the solenoid magnet as well as 
  the transition region between the CAL parts corresponding 
  to a polar angle of the positron of
  $35.6^\circ < \theta < 37.3^\circ$ and
  $128.2^\circ < \theta < 140.2^\circ$.  
\end{itemize} If more than one positron candidate in an 
event passes these cuts ($7\%$ of the events), the one with the
highest probability is assumed to be the DIS positron.
\item Elastic Compton scattering events ($e p \rightarrow e \gamma p$) are
  removed by searching for an additional photon candidate and
  discarding the event if this and the positron candidate account 
  for all the calorimeter energy in the event except for at most 2~\gev.
\item To further reduce background 
  from photoproduction, $y$ estimated from
  the positron energy and angle is required to be $y_{e} < 0.95$.
\item The net transverse momentum \Pt\ is expected to be close to zero and
  is measured with an error
  approximately proportional to $\sqrt{\Et}$. To remove cosmic rays
  and beam-related background, \Pt\ is required to be less than
  $4\sqrt{\Et\,(\gev)}$.
\end{itemize}

The efficiency of these cuts for selecting DIS events with \qqgt~400~\gevv\ 
as determined by MC is, on average, $80\%$. 
It is approximately uniform
over the kinematic phase space except for the region of high $y$
and low \qq\ where the efficiency decreases due to the positron 
energy requirement.
Uncertainties in the simulation of the efficiency arising from
the diffractive contribution to the cross-section which is not
included in the MC are neglected since the diffractive contribution
is small for \qqgt 400~\gevv.

\subsection{Final event sample}
\label{sc:selec:sample}

\begin{figure} [p]
\centerline{\epsfig{file=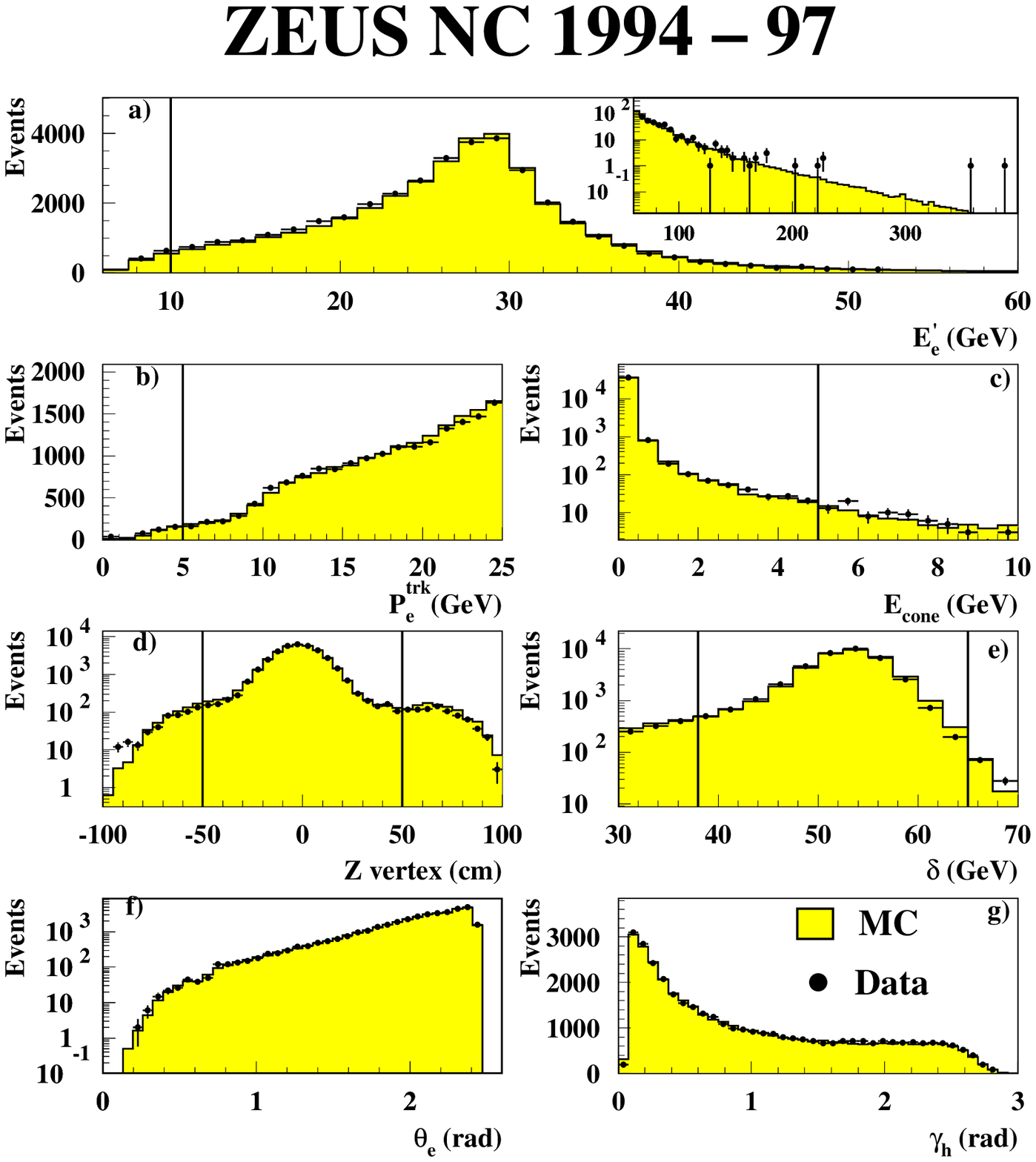,width=.9\linewidth}}
\caption[Fig1]{A comparison of data (points) and the signal MC predictions
  (histograms) for \\ (a) $E_e^\prime$, the energy of the scattered positron;
  the inset shows the high energy part of the distribution; (b)
  $P^{\rm trk}_{e}$, the momentum of the positron track; (c) the energy
  $E_{\rm cone}$
  not assigned to the positron in an $\eta$-$\Phi$ cone of 0.8; (d)
  the $Z$ position of the event vertex; (e) $\delta = \Sigma ( E_i-
  p_{zi})$; (f) $\theta_e$, the angle of the positron; and (g)
  \gh, the angle of the hadronic system. Only events which pass all 
  event selection cuts are plotted in (e) and (f);
  in (a)--(e), events are plotted which pass all other event selection cuts
  except those on the variable displayed; the cuts to be applied to this 
  variable are indicated by the vertical lines.}
\label{fg:datavsmc}
\end{figure}

After the event selection 37438 events with \qqda $>$ 400~\gevv\ remain.  
Distributions from data and signal MC are compared in Fig.~\ref{fg:datavsmc}.
Shown are the following: the positron energy; the momentum of the positron 
track; the energy not assigned to the positron in an
$\eta-\Phi$ cone of 0.8; the $Z$ position of the event vertex; the
$\delta$-distribution; the scattering angle of the positron, $\theta_{e}$; 
and \gh, the angle of the hadronic system as obtained from~(\ref{eq:gammah}). 
Good agreement between data and MC is seen in most variables. Slight
disagreements between data and MC at lower positron energies and at
low values of \gh\ may indicate simulation errors in either fragmentation 
or detector response and are accounted for in the systematic uncertainties
(see Sect.~\ref{sc:syserr}).

The photoproduction background is estimated to be less than $0.5\%$ 
over the full phase space and less than $3\%$ at high $y$. 
Background from prompt photon events is negligible.

Backgrounds not related to $ep$ reactions, such as cosmic rays or 
beam-related background, are investigated by studying events in the 
tails of distributions of the calorimeter timing and of $\Pt/\sqrt{\Et}$, 
and events from unpaired positron or proton bunches. No evidence for 
such background is observed  and an 
upper limit of $0.05\%$ at low \qq, rising to $0.6\%$ at high \qq, is 
placed on any such contamination.

\section{Cross-section determination}
\label{sc:unfold}

The single differential cross-sections are determined using
bin-by-bin unfolding.  The measured cross-section
in a particular bin, $\sigma_{\rm meas}$, is determined from
\begin{equation}
  \sigma_{\rm meas} = \frac{N_{\rm obs} - N_{\rm bg}}{\cal{A} \cal{L}},
  \label{eq:Rad}
\end{equation}
where $N_{\rm obs}$ is the number of observed events in the bin, 
$N_{\rm bg}$ is the estimated number of background events, $\cal{A}$ is the
acceptance and $\cal{L}$ is the integrated luminosity.  The acceptance, 
defined from the MC as the number of events reconstructed within the bin
divided by the number of events generated in that bin, derives from the
selection efficiency and the purity in the bin.

The measured cross-section includes the radiative effects discussed in
Sect.~\ref{sc:standardmodel}.
The correction factor to provide the Born level cross-section is defined as
\begin{equation}
  \mathcal{C}\rad = 
  \frac{\sigma\smod\born}{\sigma\smod\rad}
  \label{eq:crad} .
\end{equation}
The numerator is obtained by numerically integrating~(\ref{eq:born}) over
the bin with fixed $\alpha = 1/137$, \Mz\ = 91.175~\gev, and \sthw = 0.232.
The value of $\sigma^{\rm SM}_{\rm rad}$, the bin cross-section 
including radiation, is calculated using the HERACLES
MC. The measured Born level cross-section is then given by
\begin{equation}
   \sigma\born = \sigma_{\rm meas} \, \mathcal{C}\rad .
\end{equation}

Finally, the quoted differential cross-section, for example \sigqq, 
is calculated as
\begin{equation}
\frac{d\sigma}{dQ^2} =
\frac{\sigma\born}{\sigma\smod\born}\,
\frac{d\sigma\smod\born}{dQ^2} .
\end{equation}
A similar procedure is used for \sigx\ and \sigy. In this manner the effects of
all the selection cuts are corrected (Sect.~\ref{sc:selec}) and the
cross-sections are extrapolated to the full kinematic range. In particular
the MC is used to extrapolate beyond the $y$-region restricted by
the $y_e$ cut.

For the differential cross-section \sigqq\ the selected events are
divided into 20 bins in ${\log}_{10} \qqda$. The first 12 of these
bins divide the \qq\ range of 400 to 3200~\gevv\ into bins of equal
width. The remaining 8 bins divide the \qq\ range of 3200 to
51200~\gevv\ using bins that are twice as wide. For \sigx, 
the events are divided in
${\log}_{10} x$ into five bins per decade
in the range $x\leq 0.4$ for \qqgt 400~\gevv, $x\leq 0.63$ for 
\qqgt 2500~\gevv\ and $x\leq 1$ for \qqgt 10000~\gevv. These limits 
restrict the analysis to a region of small bin-to-bin migration.
To measure the
differential cross-section \sigy, the selected events are divided into
$y$ bins of width 0.05 for \qqgt 400 and 2500~\gevv\ and bins twice as
large for \qqgt 10000~\gevv.  
The values at which \sigqq\ and \sigx\ are quoted,
\qqc\ and $x_c$, are chosen to be near the logarithmic center of each bin.
The cross-section \sigy\ is quoted at the center $y_c$ of each bin.

Typical bin purities are $75\%$, 
where the purity quantifies migration effects and is
defined as the ratio of the number of events generated and measured 
in the bin to the number of events measured in the bin.

For \qqgt\ 400~\gevv, the \qqda\ resolution given by the r.m.s.\ of 
$(\qqda-\qq)/\qq$ is approximately constant at $5\%$. \qqda\ is slightly 
biased towards higher \qq\ values, mainly due to the effect of initial-state 
QED radiation.

The distribution in $(\xda-x)/x$ has an r.m.s.\ of $9\%$ for
$0.03<x<0.2$.  In the remaining part of the considered $x$ range the 
r.m.s.\ increases to $12\%$. The distribution in
($\yda-y$) has an r.m.s.\ of 0.05, independent of $y$.

The statistical errors are calculated using the square root of the number
of measured events $N$ for $N>100$ and otherwise from $68\%$ Poisson confidence
intervals around $N$.

The values of \qqc, $x_c$ and $y_c$, 
the number of observed events, $N_{\rm obs}$, the estimated number of 
photoproduction background events, $N_{\rm bg}$, the acceptance, ${\cal{A}}$, 
and the radiative correction factor,
$\mathcal{C}\rad$, are given in Tables~\ref{tb:xsecq2} to~\ref{tb:xsecyt}.

\section{Systematic uncertainties}
\label{sc:syserr}

Systematic uncertainties associated with detector effects are estimated by
re-calculating the cross-section after re-weighting and modifying the
MC to account for
discrepancies between data and MC in reconstruction and
efficiencies.  Cut values are varied where this method is not
applicable. The individual uncertainties are added in quadrature 
separately for the positive 
and negative deviations from the nominal cross-section values to obtain 
the total systematic uncertainty. The uncertainty on the luminosity of the 
combined 1994\,--1997 sample is \lume\ and is not included in the total 
systematic uncertainty. The remaining uncertainties are discussed in 
detail below\footnote{Note that the limits of error referred to
are on the {\it absolute magnitude} of the error.}:

\begin{itemize}
\item The main uncertainty in the trigger chain is expected to come from the
  first level. Re-weighting the MC efficiency to that derived
  from data results in systematic uncertainties below $1\%$.
\item The positron identification efficiency predicted by the MC
  is checked with
  a data sample of NC DIS events selected using independent
  requirements such as high $E_T$ in the trigger and an
  isolated high-$p_t$ track which is associated with the
  scattered positron. The efficiency curves from MC
  and data agree to better than $0.3\%$ without a bias.  For lower
  positron energies, the efficiency is checked using elastic QED
  Compton scattering events (see Sect.~\ref{sc:selec:offline}). The 
  difference in the
  efficiencies between data and MC is found to be smaller
  than $1.5\%$ at the smallest electron 
  energy. In addition, loose cuts for forward, high-energy positrons
  beyond the tracking acceptance are used to select candidates which
  are then inspected.  The results are consistent with the MC.
\item The uncertainty in the positron energy scale (as
  described in Sect.~\ref{sc:cperf}) results in systematic
  variations in the cross-section of $1\%$ at high $y$ and in
  negligible uncertainties otherwise.  Varying the positron
  identification efficiency according to the elastic
  QED Compton scattering
  events and the track finding efficiency, as derived from
  a comparison between data and MC, produces uncertainties of at most
  $2\%$. The positron isolation requirement is estimated  by varying the 
  cut value by $\pm$~2~\gev\ and causes systematic uncertainties of up 
  to $2\%$. Uncertainties in the measurement of the positron
  angle produce a systematic variation of up to $2\%$; not using the fiducial
  cut resulted in variations of less than $2\%$.
\item The uncertainty associated with the reconstruction of \gh\ is
  investigated as follows: 
  by varying the calorimeter energy scale separately for
  RCAL, BCAL, FCAL according to the energy scale uncertainty 
  described in Sect.~\ref{sc:cperf};
  by varying the $\gamma_{\rm max}$ parameter in the correction
  described in Sect.~\ref{sc:recogamma} in a range still compatible with an
  optimal reconstruction of \gh; by changing the energy of the
  calorimeter cells adjacent to the forward beampipe (and not associated
  with the current jet) based on the uncertainty estimated from a data-MC 
  comparison
  reflecting uncertainties in the simulation of the
  proton remnant; by excluding events with \gh$<0.1$~rad to check for
  the effect of loss of hadronic energy through the forward beam hole of
  the calorimeter; by not using the modified hadronic energy resolution 
  in the MC (see Sect.~\ref{sc:cperf}); 
  and by exploring the differences between
  predictions from the LEPTO~\cite{meps} and ARIADNE~\cite{ariadne}
  models of fragmentation. The last mentioned effect gives the dominating
  contribution to the systematic uncertainty. The net result is an estimated
  systematic uncertainty of less than $3\%$ in the differential cross-sections
  at low \qq\ and low $x$, increasing to approximately $8\%$ 
  at high \qq\ or high $x$.
\item The uncertainty arising from the limited knowledge of the shape of
  the vertex distribution in 
  the $Z$ coordinate (see Sect.~\ref{sc:mc}) is at most $1\%$.
\item Systematic uncertainties due to background removal are estimated by
  varying the cuts on $\delta$ and $y_e$ in a range that
  changes the expected background by more than $10\%$
  and varying the cut on $\Pt/\sqrt{\Et}$ such that signal events are strongly 
  affected.  The uncertainties in the cross-section are below $2\%$ 
  for most of the kinematic range; they increase to $8\%$
  at high \qq\ due to the $y_e$ cut.
  The systematic uncertainty arising from a
  possible underestimation of the photoproduction background is obtained from
  the effect of doubling the background predicted from the MC.
  This results in negligible changes in the cross-sections for most of the 
  kinematic range, except at high $y$ where a change of $3\%$ is found.
\end{itemize}

An important cross check is the determination of the cross-section
using the positron variables~\cite{recmethod} rather than the double-angle 
variables. The results from the two methods agree to better 
than $2\%$ for all points.

\section{Results}

\begin{figure} [p]
\centerline{\epsfig{file=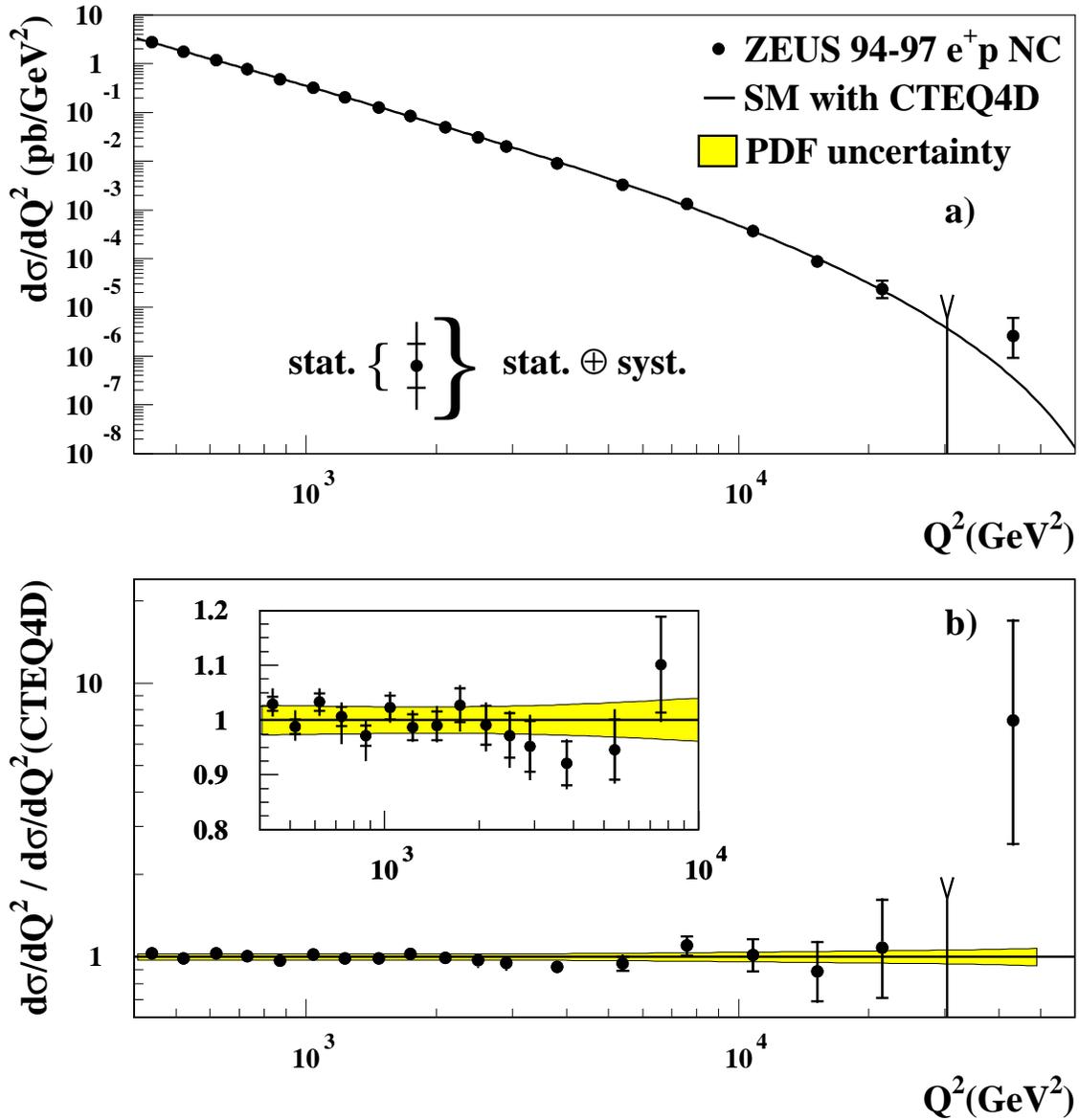,width=.9\linewidth}}
\caption[this space for rent]{The high-\qq\ \ep\ NC DIS cross-section,
  \sigqq, for data (points with error bars) and the Standard Model
  predictions using the CTEQ4D parton momentum distributions (a). Also
  plotted is the ratio of data to the prediction (b).  The
  inner error bars (delimited by the horizontal lines) show the
  statistical errors, the outer ones the statistical and
  systematic uncertainties added in quadrature.  
  The shaded region gives the uncertainty in the
  Standard Model prediction due to the uncertainty in the
  parton momentum distributions (PDF)~\cite{fitMichiel}.}
\label{fg:xsecq2}
\end{figure}

\begin{figure} [p]
\centerline{\epsfig{file=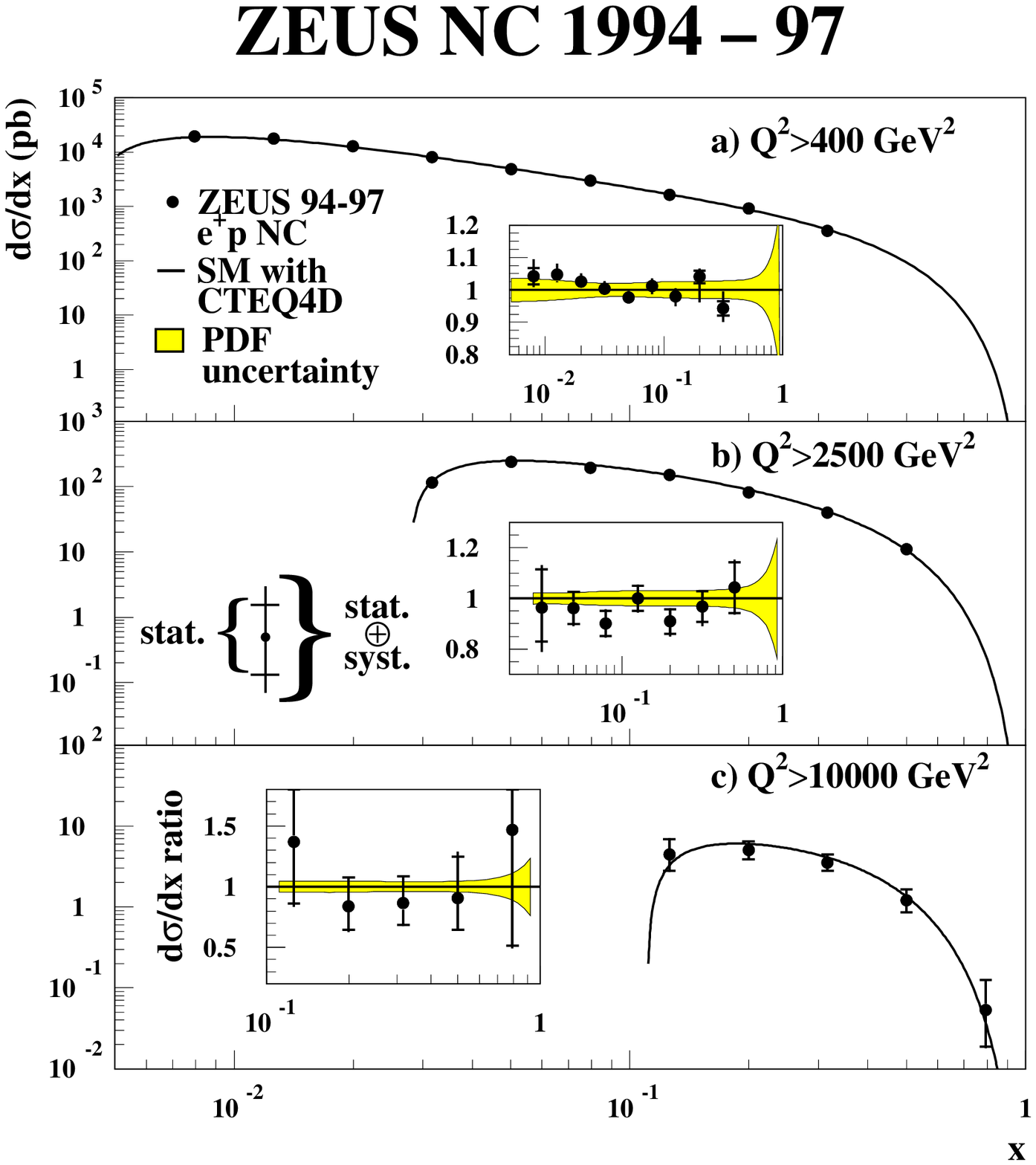,width=.9\linewidth}}
\caption[this space for rent]{The high-\qq\ \ep\ NC DIS cross-section,
  \sigx, for data (points with error bars) and the Standard Model
  predictions using the CTEQ4D parton momentum distributions.  Plotted
  are cross-sections for (a) \qqgt 400~\gevv, (b) \qqgt 2500~\gevv,
  and (c) \qqgt 10000~\gevv.  The
  inner error bars (delimited by the horizontal lines) show the
  statistical errors, the outer ones the statistical and
  systematic uncertainties added in quadrature. The shaded
  region gives the uncertainty in the Standard Model prediction due to
  the uncertainty in the parton momentum distributions (PDF)~\cite{fitMichiel}.
  The insets show the ratio of data to the prediction.}
\label{fg:xsecx}
\end{figure}

\begin{figure} [p]
\centerline{\epsfig{file=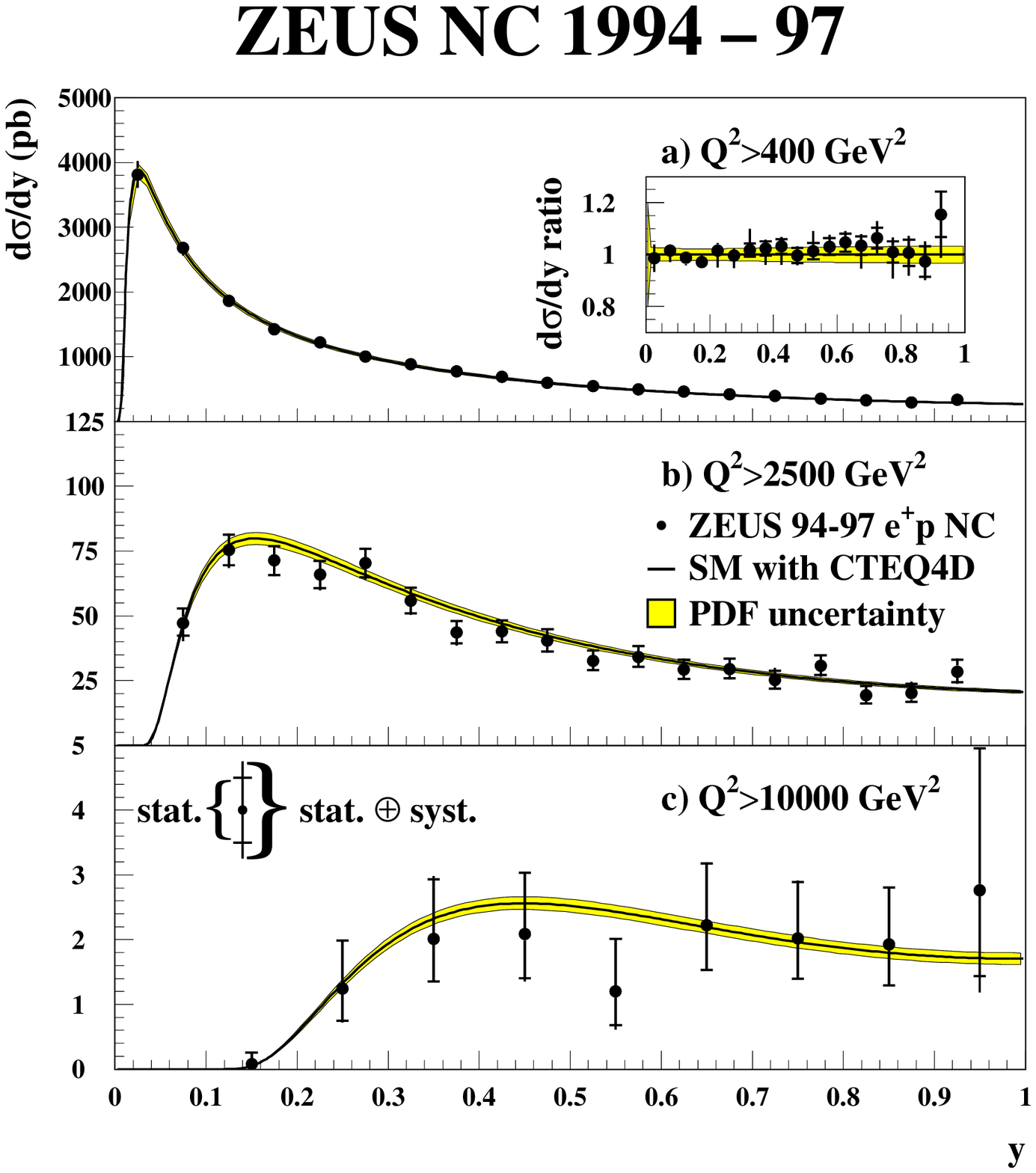,width=.9\linewidth}}
\caption[this space for rent]{The high-\qq\ \ep\ NC DIS cross-section,
  \sigy, for data (points with error bars) and the Standard Model
  predictions using the CTEQ4D parton momentum distributions.  Plotted
  are cross-sections for (a) \qqgt 400~\gevv, (b) \qqgt 2500~\gevv,
  and (c) \qqgt 10000~\gevv.  The
  inner error bars (delimited by the horizontal lines) show the
  statistical errors, the outer ones the statistical and
  systematic uncertainties added in quadrature. The shaded
  region gives the uncertainty in the Standard Model prediction due to
  the uncertainty in the parton momentum distributions (PDF)~\cite{fitMichiel}.
  The inset in (a) shows the ratio of data to the prediction.}
\label{fg:xsecy}
\end{figure}

The differential cross-sections for NC scattering, \sigqq, 
\sigx\ and \sigy\ are presented in Figs.~\ref{fg:xsecq2}
to~\ref{fg:xsecy} and Tables~\ref{tb:xsecq2} to~\ref{tb:xsecyt} 
as functions of \qq, $x$ and $y$, respectively. 
The cross-section \sigqq\ decreases by six orders of magnitude 
between \qq~=~400 and 40000~\gevv. This decrease is dominated by 
the photon propagator. The cross-section \sigx\ is shown for 
different \qq\ regions, \qq\ above 400, 2500 and 10000 GeV$^2$, 
respectively. A slow fall-off is observed towards $x=0.5$ followed 
by a rapid drop towards $x=1$.  The selection \qqgt 10000~\gevv\ limits 
the NC process by kinematics to the region $x > 0.1$ where 
the contribution from valence quarks ($u_v,d_v$) is expected to dominate. 
The cross-section \sigy\ is presented for the same regions in \qq\ as 
used for \sigx. 
For \qqgt 400~\gevv\ the bulk of the cross-section is concentrated at 
small values of $y$. For \qqgt 10000~\gevv\ the cross-section is 
approximately constant with $y$.

The predictions of the Standard Model (solid curves with PDF uncertainties,
see Sect.~\ref{sc:standardmodel}) 
give a good description of all measured cross-sections, except 
for \sigqq\ in the highest bin with \qqgt 36200~\gevv\ where two
events are observed while 0.27 are predicted by the SM. These events were 
reported previously~\cite{zhighx} as part of an excess seen at high $x$ and
high $y$, 
obtained from the first half of the data. No additional events were observed 
in the high-\qq\ bin after doubling the integrated luminosity. 
As mentioned earlier, 
the present analysis is not optimized for the detection of narrow high-mass 
lepton-hadron resonances; an analysis of this type is in progress.  

\begin{figure} [p]
\centerline{\epsfig{file=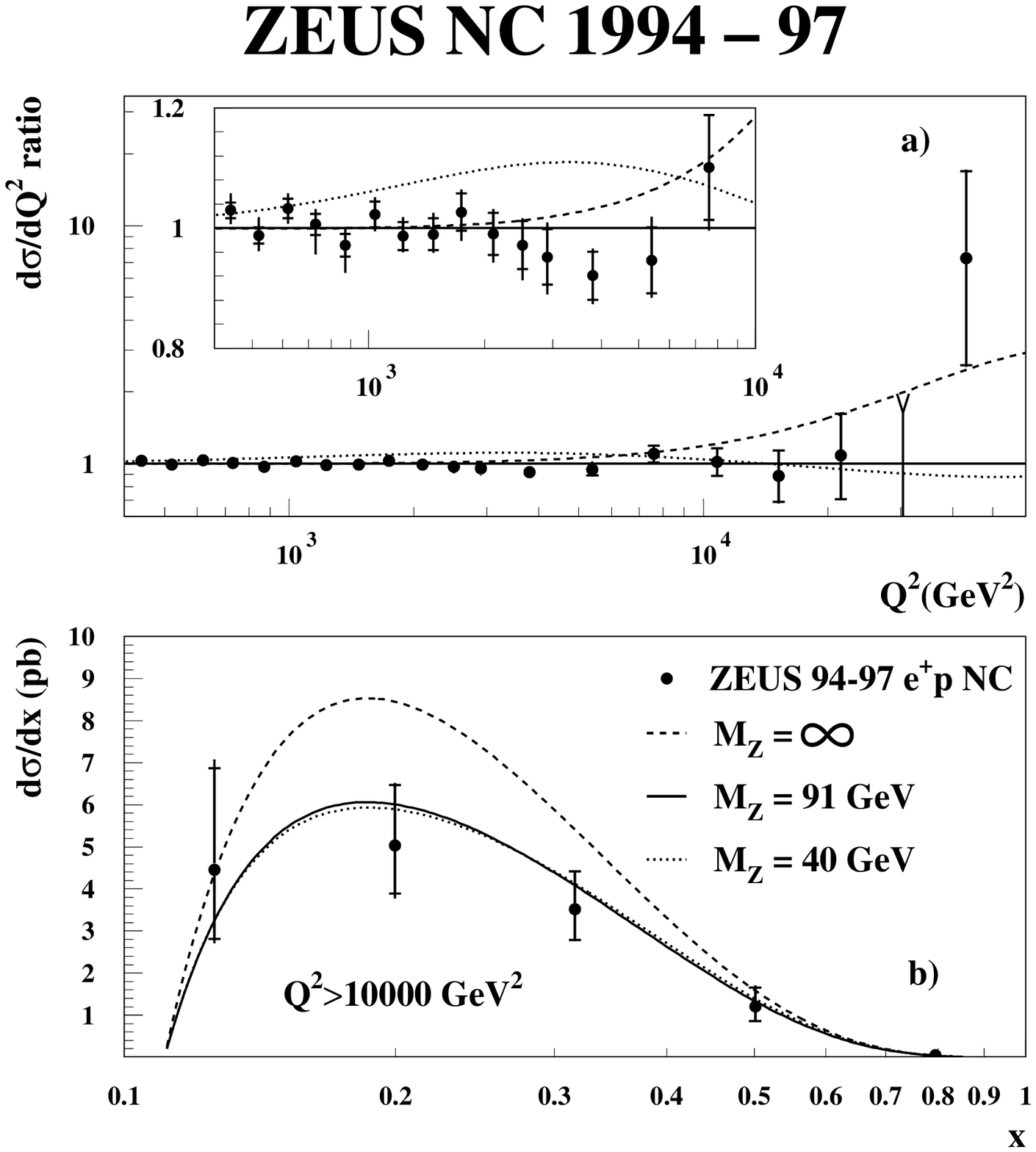,width=.9\linewidth}}
\caption[this space for rent]{The points with error bars give the ratio 
  of the measured cross-sections, \sigqq, to the SM prediction using 
  the CTEQ4D parton momentum  distributions and fixing the $Z^0$ 
  mass, \Mz, at 
  its nominal value of 91.175 GeV (a) and \sigx\ for \qqgt 
  10000~\gevv\ (b). 
  The three lines show the SM predictions for \Mz\ = 91.175~GeV 
  (solid line), for \Mz\ = 40~GeV (dotted line) and for no 
  $Z^0$ contribution (dashed line).
  The inner error bars (delimited by the horizontal lines) show the
  statistical errors, the outer ones the statistical and
  systematic uncertainties added in quadrature.}
\label{fg:xsecz}
\end{figure}

NC scattering at high \qq\ is sensitive to the contribution from the 
$Z^0$. 
According to the SM, the $Z^0$ contribution reduces 
the cross-section for \qqgt 10000~\gevv\ by about $25\%$. 
The presence of the $Z^0$ contribution 
in NC deep inelastic scattering was first demonstrated at \qq\ around 
1-2~\gevv\ through the observation of an asymmetry in the scattering of 
polarized electrons on deuterons~\cite{Prescott}. The high precision data 
from the present analysis provide sensitivity in the inclusive NC DIS 
cross-section to the $Z^0$ contribution at high \qq.
This is of particular interest as this measurement in the space-like region 
is complementary to the time-like production of the $Z^0$ 
in $p\bar p$ and 
$e^+e^-$ annihilation and thus is an important test of the Standard Model. In 
Fig.~\ref{fg:xsecz} the measured cross-sections are compared with the SM 
predictions, varying the mass of $Z^0$ in the propagator 
(see~(\ref{eq:chi})), to values of $M_Z = 40,\; 91$ GeV and infinity, 
while keeping the couplings fixed. 
Figure~\ref{fg:xsecz}(a) shows the ratio of the measured cross-section 
\sigqq\ to the prediction of the SM and Fig.~\ref{fg:xsecz}(b) 
presents \sigx\ as a function of $x$ for \qqgt 10000~\gevv. 
After separately normalizing the SM prediction with or without $Z^0$ exchange
to the data in the full \qq\ range available (and thus
essentially eliminating uncertainties arising from either the luminosity
measurement or the PDFs), $\chi^2$ values are calculated from the 
cross-sections in a \qq\ range sensitive to $Z^0$-exchange (\sigqq\ for 
$2100<$\qq$<10000$~\gevv\ and \sigx\ for \qqgt 10000~\gevv). 
Considering statistical errors only, the SM prediction
yields $\chi^2 =$ 10.3 for 10 degrees of freedom, corresponding to a
probability prob(SM)$=41\%$. In contrast, omitting the $Z^0$ 
contribution to the
cross-section, yields $\chi^2 =$ 26.4, 
i.e.\ prob(\mbox{SM without $Z^0$})$=0.3\%$. 
Taking into account each source of systematic uncertainties at a time
induces $\chi^2$ variations 
in the range 8.9-11.8 (22-29) for the SM prediction with (without) 
$Z^0$ exchange, implying prob(SM without $Z^0$)$<1.4\%$.

\section{Summary}

We have studied deep inelastic neutral-current $e^+p$ scattering based 
on data collected during 1994\,--1997 with a total luminosity of 
\lumi. The differential cross-sections \sigqq\ (for 400\qqbt 51200~\gevv) 
and \sigx, \sigy\ (for \qqgt 400, 2500, 10000~\gevv) have been measured 
with typical 
statistical and systematic errors of 3-$5\%$ for \qqlt 10000~\gevv. 
The cross-section \sigqq\ falls by six orders of magnitude 
between \qq~=~400 and 40000~\gevv. The predictions of the Standard 
Model are in very good agreement with the data. Complementing the 
observations of time-like $Z^0$ contributions to
fermion-antifermion annihilation, these data provide direct evidence for
the presence of $Z^0$ exchange in the space-like region explored by deep
inelastic scattering.

\setcounter{secnumdepth}{0} 
\section{Acknowledgments}

This measurement was made possible by the inventiveness and the diligent
efforts of the HERA machine group and the DESY computing staff.
The strong support and encouragement of the DESY directorate has been
invaluable.
The design, construction, and installation of the ZEUS detector has been made
possible by the ingenuity and dedicated effort of many people from inside DESY
and from the home institutes who are not listed as authors. Their
contributions are acknowledged with great appreciation.

This paper was completed shortly after the tragic and untimely death of
Prof.~Dr. B.~H.~Wiik, Chairman of the DESY directorate. All members of
the ZEUS collaboration wish to acknowledge the remarkable r\^ole which
he played in the success of both the HERA project and of the ZEUS
experiment. His inspired scientific leadership, his warm personality and
his friendship will be sorely missed by us all.

%
%
\clearpage
\begin{table} [!ht] \caption[this space for rent]{The differential 
cross-section \sigqq\ for the reaction $e^{+} p \rightarrow e^{+} X$.  
The following quantities are given for each bin: the \qq\ range;
the value at which the cross-section is quoted, \qqc; the number of 
selected events, $N_{\rm obs}$; the number of expected background 
events, $N_{\rm bg}$; the acceptance, $\cal A$; the radiative 
correction factor, ${\cal C}_{\rm rad}$ (see Sect.~\ref{sc:unfold});
the measured Born--level cross-section \sigqq; and the Born--level 
cross-section predicted by the Standard Model
using CTEQ4D parton momentum distributions.
The first error of each measured cross-section gives the statistical 
error, the second the systematic uncertainty.}
\label{tb:xsecq2}
\vskip 0.5 cm
\begin{center}
{\footnotesize
\renewcommand{\arraystretch}{1.2}
\begin{tabular}{|r@{ -- }r|r|r|r|c|c|l@{}l@{$\,$}l@{$\,$}l@{$\,$}l|l|}
\hline
\multicolumn{2}{|c|}{\qq\ range} &\multicolumn{1}{c|}{\qqc}& 
  \lw{$N_{\rm obs}$} & \lw{$N_{\rm bg}$} & \lw{$\cal A$} & 
  \lw{${\cal C}_{\rm rad}$} & \multicolumn{6}{c|}{\sigqq\ (pb/\gevv)}  \\ 
  \cline{8-13}\multicolumn{2}{|c|}{(\gevv)} &\multicolumn{1}{c|}{(\gevv)}
  &&&&& \multicolumn{5}{c|}{measured} & \multicolumn{1}{c|}{SM} \\ \hline
  400.0&  475.7 &   440 &  8504 &  2.4 & 0.79 & 0.94 & & $ 2.753$ & 
$\pm  0.035$ & $^{+  0.066}_{- 0.051}$ & & $ 2.673$ \\
  475.7&  565.7 &   520 &  6319 &  2.4 & 0.77 & 0.92 & & $ 1.753$ & 
$\pm  0.024$ & $^{+  0.047}_{- 0.039}$ & & $ 1.775$ \\
  565.7&  672.7 &   620 &  5008 &  2.8 & 0.76 & 0.94 & & $ 1.187$ & 
$\pm  0.018$ & $^{+  0.022}_{- 0.023}$ & & $ 1.149$ \\
  672.7&  800.0 &   730 &  3951 &  3.2 & 0.80 & 0.94 & $($ & $7.71$ & 
$\pm  0.13$ & $^{+  0.14}_{- 0.36})$ & $\cdot 10^{-1}$ & 
$ 7.65 \cdot 10^{-1}$ \\
  800.0&  951.4 &   870 &  3210 &  6.7 & 0.87 & 0.93 & $($ & $4.79$ & 
$\pm  0.09$ & $^{+  0.10}_{- 0.21})$ & $\cdot 10^{-1}$ & 
$ 4.93 \cdot 10^{-1}$ \\
  951.4& 1131.4 &  1040 &  2641 &  3.2 & 0.89 & 0.94 & $($ & $3.21$ & 
$\pm  0.07$ & $^{+  0.06}_{- 0.06})$ & $\cdot 10^{-1}$ & 
$ 3.13 \cdot 10^{-1}$ \\
 1131.4& 1345.4 &  1230 &  2000 &  1.2 & 0.90 & 0.92 & $($ & $2.01$ & 
$\pm  0.05$ & $^{+  0.04}_{- 0.03})$ & $\cdot 10^{-1}$ & 
$ 2.04 \cdot 10^{-1}$ \\
 1345.4& 1600.0 &  1470 &  1531 &  1.6 & 0.91 & 0.93 & $($ & $1.27$ & 
$\pm  0.03$ & $^{+  0.03}_{- 0.02})$ & $\cdot 10^{-1}$ & 
$ 1.28 \cdot 10^{-1}$ \\
 1600.0& 1902.7 &  1740 &  1204 &  2.0 & 0.91 & 0.92 & $($ & $8.49$ & 
$\pm  0.26$ & $^{+  0.17}_{- 0.30})$ & $\cdot 10^{-2}$ & 
$ 8.26 \cdot 10^{-2}$ \\
 1902.7& 2262.8 &  2100 &   863 &  0.4 & 0.91 & 0.93 & $($ & $4.97$ & 
$\pm  0.18$ & $^{+  0.11}_{- 0.16})$ & $\cdot 10^{-2}$ & 
$ 5.01 \cdot 10^{-2}$ \\
 2262.8& 2690.9 &  2500 &   629 &  0.4 & 0.92 & 0.94 & $($ & $3.05$ & 
$\pm  0.13$ & $^{+  0.06}_{- 0.14})$ & $\cdot 10^{-2}$ & 
$ 3.13 \cdot 10^{-2}$ \\
 2690.9& 3200.0 &  2900 &   455 &  1.2 & 0.91 & 0.94 & $($ & $1.99$ & 
$\pm  0.10$ & $^{+  0.07}_{- 0.09})$ & $\cdot 10^{-2}$ & 
$ 2.09 \cdot 10^{-2}$ \\
 3200.0& 4525.5 &  3800 &   565 &  3.1 & 0.91 & 0.93 & $($ & $9.00$ & 
$\pm  0.39$ & $^{+  0.20}_{- 0.24})$ & $\cdot 10^{-3}$ & 
$ 9.77 \cdot 10^{-3}$ \\
 4525.5& 6400.0 &  5400 &   303 &  0.0 & 0.91 & 0.91 & $($ & $3.30$ & 
$\pm  0.19$ & $^{+  0.17}_{- 0.10})$ & $\cdot 10^{-3}$ & 
$ 3.49 \cdot 10^{-3}$ \\
 6400.0& 9050.0 &  7600 &   162 &  0.0 & 0.90 & 0.93 & $($ & $1.32$ & 
$\pm  0.10$ & $^{+  0.02}_{- 0.07})$ & $\cdot 10^{-3}$ & 
$ 1.20 \cdot 10^{-3}$ \\
 9050.0&12800.0 & 10800 &    63 &  0.0 & 0.86 & 0.93 & $($ & $3.69$ & 
$^{+  0.53}_{- 0.47}$ & $^{+  0.08}_{- 0.11})$ & $\cdot 10^{-4}$ & 
$ 3.64 \cdot 10^{-4}$  \\
12800.0&18102.0 & 15200 &    20 &  0.0 & 0.81 & 0.93 & $($ & $8.9$ & 
$^{+  2.5}_{- 2.0}$ & $^{+  0.7}_{- 0.6})$ & $\cdot 10^{-5}$ & 
$ 10.0 \cdot 10^{-5}$  \\
18102.0&25600.0 & 21500 &     8 &  0.0 & 0.86 & 0.96 & $($ & $2.4$ & 
$^{+  1.2}_{- 0.8}$ & $^{+  0.4}_{- 0.1})$ & $\cdot 10^{-5}$ & 
$ 2.2 \cdot 10^{-5}$  \\
25600.0&36203.0 & 30400 &     0 &  0.0 & 0.86 & 0.90 & 
\multicolumn{5}{c|}{$< 6.0\cdot 10^{-6}$} & $ 3.7 \cdot 10^{-6}$ \\
36203.0&51200.0 & 43100 &     2 &  0.0 & 0.89 & 1.00 & $($ & $2.6$ & 
$^{+  3.5}_{- 1.7}$ & $^{+  0.7}_{- 0.2})$ & $\cdot 10^{-6}$ & 
$ 0.4 \cdot 10^{-6}$  \\
\hline
\end{tabular}
}
\end{center}
\end{table}

\clearpage
\begin{table} [!ht] \caption[this space for rent]{The differential 
cross-section \sigx\ for the reaction $e^{+} p \rightarrow e^{+} X$
for \qqgt 400, 2500, and 10000~\gevv.  
The following quantities are given for each bin: the \qq\ and $x$ range;
the value at which the cross-section is quoted, $x_c$; the number of 
selected events, $N_{\rm obs}$; the number of expected background 
events, $N_{\rm bg}$; the acceptance, $\cal A$; the radiative 
correction factor, ${\cal C}_{\rm rad}$ (see Sect.~\ref{sc:unfold});
the measured Born--level cross-section \sigx; and the Born--level 
cross-section predicted by the Standard Model
using CTEQ4D parton momentum distributions.
The first error of each measured cross-section gives the statistical 
error, the second the systematic uncertainty.}
\label{tb:xsecx}
\vskip 0.5 cm
\begin{center}
{\footnotesize
\renewcommand{\arraystretch}{1.2}
\begin{tabular}{|c|c@{$\,$}c|l|r|r|c|c|l@{}l@{$\,$}l@{$\,$}l@{$\,$}l|l|}
\hline
$Q^2_{\rm min}$ & \multicolumn{2}{c|}{\lw{$x$ range}} & 
  \multicolumn{1}{c|}{\lw{$x_c$}}& \lw{$N_{\rm obs}$} & 
  \lw{$N_{\rm bg}$} & \lw{$\cal A$} & \lw{${\cal C}_{\rm rad}$} & 
  \multicolumn{6}{c|}{\sigx\ (pb)}  \\ \cline{9-14}
(\gevv) &\multicolumn{2}{|c|}{ } &&&&&& \multicolumn{5}{c|}{measured} & 
  \multicolumn{1}{c|}{SM} \\ \hline
  400
&(0.63 -- $1.00)$&$\cdot 10^{-2}$ & 
0.00794 & 
 2307 &  6.3 & 0.67 & 0.98 &$($& $ 1.96$ & $\pm  0.05$ & 
$^{+  0.09}_{- 0.05})$ & $\cdot 10^{ 4}$ & $ 1.88 \cdot 10^{ 4}$ \\
&(0.10 -- $0.16)$&$\cdot 10^{-1}$ & 
0.0126 & 
 4352 & 11.0 & 0.84 & 0.94 & $($& $1.78$ & $\pm  0.03$ & 
$^{+  0.05}_{- 0.02})$ & $\cdot 10^{ 4}$ & $ 1.70 \cdot 10^{ 4}$ \\
&(0.16 -- $0.25)$&$\cdot 10^{-1}$ & 
0.0200 & 
 5026 &  6.7 & 0.85 & 0.94 & $($& $1.27$ & $\pm  0.02$ & 
$^{+  0.03}_{- 0.01})$ & $\cdot 10^{ 4}$ & $ 1.24 \cdot 10^{ 4}$ \\
&(0.25 -- $0.40)$&$\cdot 10^{-1}$ & 
0.0316 & 
 5283 &  2.7 & 0.87 & 0.92 & $($& $8.08$ & $\pm  0.12$ & 
$^{+  0.14}_{- 0.10})$ & $\cdot 10^{ 3}$ & $ 8.04 \cdot 10^{ 3}$ \\
&(0.40 -- $0.63)$&$\cdot 10^{-1}$ & 
0.0501 & 
 5028 &  2.7 & 0.87 & 0.92 & $($& $4.83$ & $\pm  0.07$ & 
$^{+  0.08}_{- 0.05})$ & $\cdot 10^{ 3}$ & $ 4.95 \cdot 10^{ 3}$ \\
&(0.63 -- $1.00)$&$\cdot 10^{-1}$ & 
0.0794 & 
 4782 &  0.0 & 0.86 & 0.93 & $($& $2.96$ & $\pm  0.05$ & 
$^{+  0.05}_{- 0.06})$ & $\cdot 10^{ 3}$ & $ 2.92 \cdot 10^{ 3}$ \\
&0.10 -- 0.16& & 
0.126 & 
 4219 &  0.4 & 0.86 & 0.92 & $($& $1.63$ & $\pm  0.03$ & 
$^{+  0.03}_{- 0.04})$ & $\cdot 10^{ 3}$ & $ 1.66 \cdot 10^{ 3}$ \\
&0.16 -- 0.25& & 
0.200 & 
 3512 &  0.0 & 0.82 & 0.92 & $($& $9.04$ & $\pm  0.17$ & 
$^{+  0.15}_{- 0.67})$ & $\cdot 10^{ 2}$ & $ 8.68 \cdot 10^{ 2}$ \\
&0.25 -- 0.40& & 
0.316 & 
 2276 &  0.4 & 0.87 & 0.92 & $($& $3.51$ & $\pm  0.08$ & 
$^{+  0.18}_{- 0.13})$ & $\cdot 10^{ 2}$ & $ 3.72 \cdot 10^{ 2}$ \\
\hline
 2500
&(0.25 -- $0.40)$&$\cdot 10^{-1}$ & 
0.0316 & 
   58 &  1.2 & 0.72 & 1.01 & $($& $1.15$ & $^{+  0.18}_{- 0.16}$ & 
$^{+  0.09}_{- 0.13})$ & $\cdot 10^{ 2}$ & $ 1.19\cdot 10^{ 2}$ \\
&(0.40 -- $0.63)$&$\cdot 10^{-1}$ & 
0.0501 & 
  252 &  2.7 & 0.91 & 0.95 & $($& $2.40$ & $\pm  0.16$ & 
$^{+  0.06}_{- 0.09})$ & $\cdot 10^{ 2}$ & $ 2.49 \cdot 10^{ 2}$ \\
&(0.63 -- $1.00)$&$\cdot 10^{-1}$ & 
0.0794 & 
  340 &  0.0 & 0.94 & 0.93 & $($& $1.94$ & $\pm  0.11$ & 
$^{+  0.06}_{- 0.05})$ & $\cdot 10^{ 2}$ & $ 2.16 \cdot 10^{ 2}$ \\
&0.10 -- 0.16& & 
0.126 & 
  421 &  0.0 & 0.93 & 0.91 & $($& $1.51$ & $\pm  0.08$ & 
$^{+  0.03}_{- 0.04})$ & $\cdot 10^{ 2}$ & $ 1.51 \cdot 10^{ 2}$ \\
&0.16 -- 0.25& & 
0.200 & 
  356 &  0.0 & 0.93 & 0.92 & $($& $8.16$ & $\pm  0.44$ & 
$^{+  0.16}_{- 0.31})$ & $\cdot 10^{ 1}$ & $ 8.97 \cdot 10^{ 1}$ \\
&0.25 -- 0.40& & 
0.316 & 
  265 &  0.4 & 0.87 & 0.91 & $($& $4.02$ & $\pm  0.25$ & 
$^{+  0.08}_{- 0.20})$ & $\cdot 10^{ 1}$ & $ 4.15 \cdot 10^{ 1}$ \\
&0.40 -- 0.63& & 
0.501 & 
  112 &  0.0 & 0.84 & 0.93 & $($& $1.10$ & $\pm  0.11$ & 
$^{+  0.05}_{- 0.04})$ & $\cdot 10^{ 1}$ & $ 1.06 \cdot 10^{ 1}$ \\
\hline
10000
&0.10 -- 0.16& & 
0.126 & 
    7 &  0.0 & 0.62 & 1.06 & & $ 4.5$ & $^{+  2.4}_{- 1.7} $ & 
$^{+  1.0}_{- 0.6}$ & & 3.3 \\
&0.16 -- 0.25& & 
0.200 & 
   19 &  0.0 & 0.82 & 0.94 & & $ 5.0$ & $^{+  1.4}_{- 1.2} $ & 
$^{+  0.4}_{- 0.5}$ & & 6.0 \\
&0.25 -- 0.40& & 
0.316 & 
   23 &  0.0 & 0.88 & 0.91 & & $ 3.5$ & $^{+  0.9}_{- 0.7} $ & 
$^{+  0.1}_{- 0.2}$ & & 4.1 \\
&0.40 -- 0.63& & 
0.501 & 
   12 &  0.0 & 0.85 & 0.93 & & $ 1.2$ & $^{+  0.5}_{- 0.3} $ & 
$^{+  0.2}_{- 0.1}$ & & 1.3 \\
&0.63 -- 1.00& & 
0.794 & 
    2 &  0.0 & 0.92 & 0.95 & $($& $5.4$ & $^{+  7.1}_{- 3.5}$ & 
$^{+  2.1}_{- 0.8})$ & $\cdot 10^{-2}$ & $ 3.6\cdot 10^{-2}$ \\
\hline
\end{tabular}
}
\end{center}
\end{table}

\clearpage
\begin{table} [!ht] \caption[this space for rent]{The differential 
cross-section \sigy\ for the reaction $e^{+} p \rightarrow e^{+} X$
for \qqgt 400~\gevv.  
The following quantities are given for each bin: the \qq\ and $y$ range;
the value at which the cross-section is quoted, $y_c$; the number of 
selected events, $N_{\rm obs}$; the number of expected background 
events, $N_{\rm bg}$; the acceptance, $\cal A$; the radiative 
correction factor, ${\cal C}_{\rm rad}$ (see Sect.~\ref{sc:unfold});
the measured Born--level cross-section \sigy; and the Born--level 
cross-section predicted by the Standard Model
using CTEQ4D parton momentum distributions.
The first error of each measured cross-section gives the statistical 
error, the second the systematic uncertainty.}
\label{tb:xsecyo}
\vskip 0.5 cm
\begin{center}
{\footnotesize
\renewcommand{\arraystretch}{1.2}
\begin{tabular}{|c|c|l|r|r|c|c|l@{$\,$}l@{$\,$}l@{$\,$}l|l|}
\hline
$Q^2_{\rm min}$ & \lw{$y$ range} & \multicolumn{1}{c|}{\lw{$y_c$}}& 
  \lw{$N_{\rm obs}$} & \lw{$N_{\rm bg}$} & \lw{$\cal A$} & 
  \lw{${\cal C}_{\rm rad}$} & \multicolumn{5}{c|}{\sigy\ (pb)} \\ \cline{8-12}
(\gevv) &&&&&&& \multicolumn{4}{c|}{measured} & \multicolumn{1}{c|}{SM} \\ 
\hline
  400
&0.00 -- 0.05 & 
0.025 & 
 5613 &  0.0 & 0.80 & 0.95 & $( 3.82$ & $\pm  0.06$ & $^{+  0.20}_{- 0.20})$ &
 $\cdot 10^{ 3}$ & $ 3.87 \cdot 10^{ 3}$ \\
&0.05 -- 0.10 & 
0.075 & 
 5844 &  0.4 & 0.84 & 0.94 & $( 2.68$ & $\pm  0.04$ & $^{+  0.04}_{- 0.11})$ &
 $\cdot 10^{ 3}$ & $ 2.64 \cdot 10^{ 3}$ \\
&0.10 -- 0.15 & 
0.125 & 
 4128 &  0.4 & 0.86 & 0.93 & $( 1.86$ & $\pm  0.03$ & $^{+  0.03}_{- 0.04})$ &
 $\cdot 10^{ 3}$ & $ 1.89 \cdot 10^{ 3}$ \\
&0.15 -- 0.20 & 
0.175 & 
 3231 &  0.0 & 0.88 & 0.93 & $( 1.43$ & $\pm  0.03$ & $^{+  0.03}_{- 0.01})$ &
 $\cdot 10^{ 3}$ & $ 1.47 \cdot 10^{ 3}$ \\
&0.20 -- 0.25 & 
0.225 & 
 2685 &  0.0 & 0.86 & 0.93 & $( 1.22$ & $\pm  0.03$ & $^{+  0.02}_{- 0.07})$ &
 $\cdot 10^{ 3}$ & $ 1.20 \cdot 10^{ 3}$ \\
&0.25 -- 0.30 & 
0.275 & 
 2226 &  0.0 & 0.85 & 0.92 & $( 1.01$ & $\pm  0.02$ & $^{+  0.02}_{- 0.04})$ &
 $\cdot 10^{ 3}$ & $ 1.01 \cdot 10^{ 3}$ \\
&0.30 -- 0.35 & 
0.325 & 
 1939 &  0.0 & 0.85 & 0.92 & $( 8.83$ & $\pm  0.22$ & $^{+  0.68}_{- 0.07})$ &
 $\cdot 10^{ 2}$ & $ 8.68 \cdot 10^{ 2}$ \\
&0.35 -- 0.40 & 
0.375 & 
 1731 &  2.4 & 0.87 & 0.93 & $( 7.76$ & $\pm  0.21$ & $^{+  0.16}_{- 0.42})$ &
 $\cdot 10^{ 2}$ & $ 7.57 \cdot 10^{ 2}$ \\
&0.40 -- 0.45 & 
0.425 & 
 1547 &  1.6 & 0.86 & 0.91 & $( 6.90$ & $\pm  0.19$ & $^{+  0.16}_{- 0.43})$ &
 $\cdot 10^{ 2}$ & $ 6.69 \cdot 10^{ 2}$ \\
&0.45 -- 0.50 & 
0.475 & 
 1389 &  0.0 & 0.90 & 0.92 & $( 5.95$ & $\pm  0.18$ & $^{+  0.12}_{- 0.15})$ &
 $\cdot 10^{ 2}$ & $ 5.96 \cdot 10^{ 2}$ \\
&0.50 -- 0.55 & 
0.525 & 
 1308 &  1.2 & 0.91 & 0.91 & $( 5.44$ & $\pm  0.17$ & $^{+  0.38}_{- 0.07})$ &
 $\cdot 10^{ 2}$ & $ 5.36 \cdot 10^{ 2}$ \\
&0.55 -- 0.60 & 
0.575 & 
 1205 &  3.6 & 0.91 & 0.91 & $( 5.00$ & $\pm  0.16$ & $^{+  0.15}_{- 0.07})$ &
 $\cdot 10^{ 2}$ & $ 4.85 \cdot 10^{ 2}$ \\
&0.60 -- 0.65 & 
0.625 & 
 1104 &  0.4 & 0.92 & 0.92 & $( 4.64$ & $\pm  0.15$ & $^{+  0.12}_{- 0.07})$ &
 $\cdot 10^{ 2}$ & $ 4.43 \cdot 10^{ 2}$ \\
&0.65 -- 0.70 & 
0.675 & 
  978 &  0.8 & 0.89 & 0.91 & $( 4.20$ & $\pm  0.15$ & $^{+  0.10}_{- 0.33})$ &
 $\cdot 10^{ 2}$ & $ 4.06 \cdot 10^{ 2}$ \\
&0.70 -- 0.75 & 
0.725 & 
  849 &  2.8 & 0.84 & 0.94 & $( 3.99$ & $\pm  0.15$ & $^{+  0.20}_{- 0.15})$ &
 $\cdot 10^{ 2}$ & $ 3.75 \cdot 10^{ 2}$ \\
&0.75 -- 0.80 & 
0.775 & 
  676 &  1.2 & 0.74 & 0.92 & $( 3.52$ & $\pm  0.15$ & $^{+  0.13}_{- 0.32})$ &
 $\cdot 10^{ 2}$ & $ 3.48 \cdot 10^{ 2}$ \\
&0.80 -- 0.85 & 
0.825 & 
  460 &  2.4 & 0.55 & 0.93 & $( 3.28$ & $\pm  0.16$ & $^{+  0.14}_{- 0.23})$ &
 $\cdot 10^{ 2}$ & $ 3.25 \cdot 10^{ 2}$ \\
&0.85 -- 0.90 & 
0.875 & 
  311 &  5.1 & 0.41 & 0.94 & $( 2.98$ & $\pm  0.18$ & $^{+  0.15}_{- 0.11})$ &
 $\cdot 10^{ 2}$ & $ 3.06 \cdot 10^{ 2}$ \\
&0.90 -- 0.95 & 
0.925 & 
  209 &  5.9 & 0.24 & 0.96 & $( 3.35$ & $\pm  0.25$ & $^{+  0.08}_{- 0.41})$ &
 $\cdot 10^{ 2}$ & $ 2.90 \cdot 10^{ 2}$ \\
\hline
\end{tabular}
}
\end{center}
\end{table}

\clearpage
\begin{table} [!ht]  
\caption[this space for rent]{Continuation of Table
    \ref{tb:xsecyo}: The tabulation of results for the differential
    cross-section \sigy\ for \qqgt\ 2500 and 10000~\gevv.}
\label{tb:xsecyt}
\vskip 0.5 cm
\begin{center}
{\footnotesize
\renewcommand{\arraystretch}{1.2}
\begin{tabular}{|c|c|l|r|r|c|c|l@{}l@{$\,$}l@{$\,$}l@{$\,$}l|l|}
\hline
$Q^2_{\rm min}$ & \lw{$y$ range} & \multicolumn{1}{c|}{\lw{$y_c$}}&
  \lw{$N_{\rm obs}$} & \lw{$N_{\rm bg}$} & \lw{$\cal A$} &
  \lw{${\cal C}_{\rm rad}$} & \multicolumn{6}{c|}{\sigy\ (pb)} \\ \cline{8-13}
(\gevv) &&&&&&& \multicolumn{5}{c|}{measured} & \multicolumn{1}{c|}{SM} \\  
\hline
 2500
&0.05 -- 0.10 & 
0.075 & 
   93 &  0.0 & 0.88 & 1.02 &$($& $ 4.73$ & $^{+  0.56}_{- 0.50}$ & 
$^{+  0.13}_{- 0.50})$ & $\cdot 10^{ 1}$ & $ 4.66 \cdot 10^{ 1}$ \\
&0.10 -- 0.15 & 
0.125 & 
  172 &  0.4 & 0.92 & 0.95 &$($& $ 7.55$ & $\pm  0.59$ & 
$^{+  0.23}_{- 0.35})$ & $\cdot 10^{ 1}$ & $ 7.72 \cdot 10^{ 1}$ \\
&0.15 -- 0.20 & 
0.175 & 
  171 &  0.0 & 0.93 & 0.93 &$($& $ 7.14$ & $\pm  0.56$ & 
$^{+  0.23}_{- 0.24})$ & $\cdot 10^{ 1}$ & $ 7.90 \cdot 10^{ 1}$ \\
&0.20 -- 0.25 & 
0.225 & 
  158 &  0.0 & 0.94 & 0.94 &$($& $ 6.60$ & $\pm  0.54$ & 
$^{+  0.14}_{- 0.29})$ & $\cdot 10^{ 1}$ & $ 7.31 \cdot 10^{ 1}$ \\
&0.25 -- 0.30 & 
0.275 & 
  169 &  0.0 & 0.94 & 0.93 &$($& $ 7.04$ & $\pm  0.56$ & 
$^{+  0.16}_{- 0.29})$ & $\cdot 10^{ 1}$ & $ 6.58 \cdot 10^{ 1}$ \\
&0.30 -- 0.35 & 
0.325 & 
  135 &  0.0 & 0.95 & 0.94 &$($& $ 5.59$ & $\pm  0.49$ & 
$^{+  0.24}_{- 0.24})$ & $\cdot 10^{ 1}$ & $ 5.87 \cdot 10^{ 1}$ \\
&0.35 -- 0.40 & 
0.375 & 
  108 &  0.0 & 0.94 & 0.91 &$($& $ 4.37$ & $\pm  0.43$ & 
$^{+  0.12}_{- 0.26})$ & $\cdot 10^{ 1}$ & $ 5.25 \cdot 10^{ 1}$ \\
&0.40 -- 0.45 & 
0.425 & 
  114 &  0.0 & 0.96 & 0.88 &$($& $ 4.40$ & $\pm  0.42$ & 
$^{+  0.19}_{- 0.08})$ & $\cdot 10^{ 1}$ & $ 4.70 \cdot 10^{ 1}$ \\
&0.45 -- 0.50 & 
0.475 & 
   99 &  0.0 & 0.95 & 0.92 &$($& $ 4.04$ & $^{+  0.46}_{- 0.41}$ & 
$^{+  0.07}_{- 0.18})$ & $\cdot 10^{ 1}$ & $ 4.23 \cdot 10^{ 1}$ \\
&0.50 -- 0.55 & 
0.525 & 
   85 &  0.0 & 0.97 & 0.89 &$($& $ 3.26$ & $^{+  0.40}_{- 0.36}$ & 
$^{+  0.19}_{- 0.07})$ & $\cdot 10^{ 1}$ & $ 3.82 \cdot 10^{ 1}$ \\
&0.55 -- 0.60 & 
0.575 & 
   86 &  0.0 & 0.93 & 0.88 &$($& $ 3.41$ & $^{+  0.42}_{- 0.38}$ & 
$^{+  0.08}_{- 0.17})$ & $\cdot 10^{ 1}$ & $ 3.48 \cdot 10^{ 1}$ \\
&0.60 -- 0.65 & 
0.625 & 
   72 &  0.0 & 0.93 & 0.90 &$($& $ 2.92$ & $^{+  0.40}_{- 0.35}$ & 
$^{+  0.15}_{- 0.06})$ & $\cdot 10^{ 1}$ & $ 3.18 \cdot 10^{ 1}$ \\
&0.65 -- 0.70 & 
0.675 & 
   72 &  0.0 & 0.92 & 0.89 &$($& $ 2.94$ & $^{+  0.40}_{- 0.35}$ & 
$^{+  0.13}_{- 0.15})$ & $\cdot 10^{ 1}$ & $ 2.93 \cdot 10^{ 1}$ \\
&0.70 -- 0.75 & 
0.725 & 
   62 &  0.0 & 0.94 & 0.91 &$($& $ 2.53$ & $^{+  0.37}_{- 0.33}$ & 
$^{+  0.27}_{- 0.06})$ & $\cdot 10^{ 1}$ & $ 2.72 \cdot 10^{ 1}$ \\
&0.75 -- 0.80 & 
0.775 & 
   74 &  0.0 & 0.93 & 0.92 &$($& $ 3.08$ & $^{+  0.41}_{- 0.37}$ & 
$^{+  0.08}_{- 0.21})$ & $\cdot 10^{ 1}$ & $ 2.54 \cdot 10^{ 1}$ \\
&0.80 -- 0.85 & 
0.825 & 
   44 &  1.6 & 0.85 & 0.93 &$($& $ 1.93$ & $^{+  0.36}_{- 0.31}$ & 
$^{+  0.14}_{- 0.12})$ & $\cdot 10^{ 1}$ & $ 2.39 \cdot 10^{ 1}$ \\
&0.85 -- 0.90 & 
0.875 & 
   44 &  1.2 & 0.83 & 0.92 &$($& $ 2.01$ & $^{+  0.37}_{- 0.32}$ & 
$^{+  0.26}_{- 0.31})$ & $\cdot 10^{ 1}$ & $ 2.27 \cdot 10^{ 1}$ \\
&0.90 -- 0.95 & 
0.925 & 
   55 &  1.2 & 0.74 & 0.93 &$($& $ 2.85$ & $^{+  0.46}_{- 0.41}$ & 
$^{+  0.16}_{- 0.23})$ & $\cdot 10^{ 1}$ & $ 2.17 \cdot 10^{ 1}$ \\
\hline
10000
&0.1 -- 0.2 & 
0.15 & 
    1 &  0.0 & 1.06 & 1.08 &$($& $ 7.8$ & $^{+ 18.0}_{- 6.5}$ & 
$^{+  0.6}_{- 1.9})$ & $\cdot 10^{-2}$ & $ 5.0 \cdot 10^{-2}$ \\
&0.2 -- 0.3 & 
0.25 & 
    6 &  0.0 & 0.98 & 0.95 && $ 1.2$ & $^{+  0.7}_{- 0.5} $ & 
$^{+  0.1}_{- 0.2}$ & & 1.3 \\
&0.3 -- 0.4 & 
0.35 & 
    9 &  0.0 & 0.92 & 0.97 && $ 2.0$ & $^{+  0.9}_{- 0.7} $ & 
$^{+  0.3}_{- 0.1}$ & & 2.3 \\
&0.4 -- 0.5 & 
0.45 & 
    9 &  0.0 & 0.82 & 0.90 && $ 2.1$ & $^{+  1.0}_{- 0.7} $ & 
$^{+  0.1}_{- 0.3}$ & & 2.6 \\
&0.5 -- 0.6 & 
0.55 & 
    5 &  0.0 & 0.80 & 0.92 && $ 1.2$ & $^{+  0.8}_{- 0.5} $ & 
$^{+  0.0}_{- 0.3}$ & & 2.4 \\
&0.6 -- 0.7 & 
0.65 & 
   10 &  0.0 & 0.86 & 0.91 && $ 2.2$ & $^{+  1.0}_{- 0.7} $ & 
$^{+  0.2}_{- 0.0}$ & & 2.2 \\
&0.7 -- 0.8 & 
0.75 & 
   10 &  0.0 & 0.95 & 0.92 && $ 2.0$ & $^{+  0.9}_{- 0.6} $ & 
$^{+  0.2}_{- 0.1}$ & & 2.0 \\
&0.8 -- 0.9 & 
0.85 & 
    9 &  0.0 & 0.93 & 0.95 && $ 1.9$ & $^{+  0.9}_{- 0.6} $ & 
$^{+  0.1}_{- 0.1}$ & & 1.8 \\
&0.9 -- 1.0 & 
0.95 & 
    4 &  0.0 & 0.31 & 1.01 && $ 2.8$ & $^{+  2.2}_{- 1.3} $ & 
$^{+ 0.1}_{- 0.9}$ & & 1.7 \\
\hline
\end{tabular}
}
\end{center}
\end{table}
%
%
%
%
\end{document}